\documentclass[aps,prd,twocolumn,superscriptaddress,nofootinbib,10pt]{revtex4-1}

\pdfoutput=1
\usepackage[utf8]{inputenc}
\usepackage{graphicx,amsmath,amsfonts,amssymb,aas_macros,slashed}
\usepackage{bbold}
\usepackage{datetime}
\usepackage{numprint}
\usepackage{enumitem}
 
\usepackage{graphicx,color,amsmath}
\usepackage{hyperref}

\allowdisplaybreaks

\newcommand{\cm}{\ensuremath{\mathrm{cm}}}

\newcommand{\eV}{\ensuremath{\mathrm{eV}}}
\newcommand{\keV}{\ensuremath{\mathrm{keV}}}
\newcommand{\MeV}{\ensuremath{\mathrm{MeV}}}
\newcommand{\GeV}{\ensuremath{\mathrm{GeV}}}
\newcommand{\TeV}{\ensuremath{\mathrm{TeV}}}

\newcommand{\me}[3]{\ensuremath \langle #1 | #2 | #3 \rangle   }

\DeclareMathOperator{\real}{Re}
\DeclareMathOperator{\imag}{Im}

\usepackage{pgfplots}
\usepackage{bm}
\usepackage{mathtools}

\newcommand{\sdark}{s_{\chi\bar\chi}} %

\usetikzlibrary{positioning,arrows,patterns}
\usetikzlibrary{decorations.markings}
\usetikzlibrary{calc}

\tikzset{
    photon/.style={decorate, decoration={snake}, draw=black},
    vector/.style={decorate, decoration={snake}, draw},
	provector/.style={decorate, decoration={snake,amplitude=2.5pt}, draw},
	antivector/.style={decorate, decoration={snake,amplitude=-2.5pt}, draw},
    fermion/.style={draw=black, postaction={decorate},
        decoration={markings,mark=at position .55 with {\arrow[draw=black]{>}}}},
    fermionbar/.style={draw=black, postaction={decorate},
        decoration={markings,mark=at position .55 with {\arrow[draw=black]{<}}}},
    fermionnoarrow/.style={draw=black},
    gluon/.style={decorate, draw=black,
        decoration={coil,amplitude=4pt, segment length=5pt}},
    scalar/.style={dashed,draw=black, postaction={decorate},
        decoration={markings,mark=at position .55 with {\arrow[draw=black]{>}}}},
    scalarbar/.style={dashed,draw=black, postaction={decorate},
        decoration={markings,mark=at position .55 with {\arrow[draw=black]{<}}}},
    scalartwo/.style={dotted,draw=black, postaction={decorate},
        decoration={markings,mark=at position .55 with {\arrow[draw=black]{>}}}},
    scalartwobar/.style={dotted,draw=black, postaction={decorate},
        decoration={markings,mark=at position .55 with {\arrow[draw=black]{<}}}},
    scalarnoarrow/.style={dashed,draw=black},
    electron/.style={draw=black, postaction={decorate},
        decoration={markings,mark=at position .55 with {\arrow[draw=black]{>}}}},
	bigvector/.style={decorate, decoration={snake,amplitude=4pt}, draw},
    vertex/.style={draw,shape=circle,fill=black,minimum size=1pt,inner sep=0pt},
    fermion2/.style={double, draw=black, postaction={decorate},
		decoration={markings,mark=at position .55 with {\arrow[draw=black]{>}}}},
    momentum/.style={draw=black,line width=0.15mm, postaction={decorate},
        decoration={markings,mark=at position 1 with {\arrow[draw=black]{>}}}}
}

\begin{document}

\title{Stellar probes of dark sector-photon interactions}

\author{Xiaoyong Chu}
\email{xiaoyong.chu@oeaw.ac.at}
\affiliation{Institute of High Energy Physics, Austrian Academy of Sciences, Nikolsdorfergasse 18, 1050 Vienna, Austria}
\author{Jui-Lin Kuo}
\email{jui-lin.kuo@oeaw.ac.at}
\affiliation{Institute of High Energy Physics, Austrian Academy of Sciences, Nikolsdorfergasse 18, 1050 Vienna, Austria}
\author{Josef Pradler}
\email{josef.pradler@oeaw.ac.at}
\affiliation{Institute of High Energy Physics, Austrian Academy of Sciences, Nikolsdorfergasse 18, 1050 Vienna, Austria}
\author{Lukas Semmelrock}
\email{lukas.semmelrock@oeaw.ac.at}
\affiliation{Institute of High Energy Physics, Austrian Academy of Sciences, Nikolsdorfergasse 18, 1050 Vienna, Austria}

\begin{abstract}
  Electromagnetically neutral dark sector particles may directly
  couple to the photon through higher dimensional effective
  operators. Considering electric and magnetic dipole moment, anapole
  moment, and charge radius interactions, we derive constraints from
  stellar energy loss in the Sun, horizontal branch and red giant
  stars, as well as from cooling of the proto-neutron star of SN1987A.
  We provide the exact formula for in-medium photon-mediated pair
  production to leading order in the dark coupling, and compute the
  energy loss rates explicitly for the most important processes,
  including a careful discussion on resonances and potential double
  counting between the processes.  Stringent limits for dark 
  states with masses below $3\,\keV$ ($40\,\MeV$) arise from red giant stars (SN1987A), implying an
  effective lower mass-scale of approximately $10^9\,\GeV$
  ($10^7\,\GeV$) for mass-dimension five, and $100 \,\GeV$
  ($2.5\,\TeV$) for mass-dimension six operators   as long as dark
  states stream freely; for the proto-neutron star, the trapping of
  dark states is also evaluated.
  Together with direct limits previously derived by us in Chu et
  al.~(2018), this provides the first comprehensive overview of the
  viability of effective electromagnetic dark-state interactions below
  the GeV mass-scale.
\end{abstract}

\maketitle

\section{Introduction}
\label{sec:introduction}

The prospect that new physics might be hiding under our noses in form
of light dark states that have been in kinematic reach for decades is
most intriguing if not seemingly preposterous.  In fact, cases exist
where new interactions are of comparable strength to the ones
encountered in the Standard Model (SM), while being compatible with
all to-date searches. The direct test of such physics,
\textit{i.e.}~new particles and interactions below the GeV-scale has
become a major field in recent
years~\cite{Essig:2013lka,Battaglieri:2017aum}, and provides a
complementary direction to the beyond-SM searches at the energy frontier.

Whereas the GeV-mass scale might comprise somewhat of a ``blind-spot''
that allows for the existence of new physics with appreciable
interactions to the SM, once the mass enters the keV-regime the landscape
changes fundamentally. Astrophysical constraints on long-lived dark
states that are derived from stellar cooling
arguments~\cite{Raffelt:1996wa} are typically so severe that the cases
for laboratory detection drastically diminish. Of course, the
observable signatures of dark states depend on the nature of the
coupling to the SM. For example, a new force can be mediated by new
scalar or vector particles. Benchmark models are then derived based on
minimality of the SM extension and on the dimensionality of the
interaction operator, and within this framework the interplay between
cosmological and astrophysical implications and direct tests is
fleshed out.

A prominent example is the vector portal, where the low-energy
phenomenology is determined by the kinetic mixing strength $\epsilon $
of the ``dark photon'' $V$ with the SM photon~\cite{Holdom:1985ag}.  If the mass of $V$,
$m_V$, originates from a Higgs mechanism, implying an additional scalar
particle in the vicinity of $m_V$, stellar cooling constraints
obliterate any prospects of probing such model below the keV-region,
as limits on millicharged particles apply.
However, a decoupling of stellar constraints as
$\epsilon^2 m_V^2$~\cite{An:2013yfc} when $m_V$ arises from a
Stuckelberg mechanism, opens the opportunity to explore a vast
parameter region through direct, laboratory searches, in particular if
$V$ is the dark matter (DM), see,
\textit{e.g.}~\cite{Pospelov:2008jk, Redondo:2008ec, Redondo:2013lna, An:2013yua,An:2014twa, Chaudhuri:2014dla, Chaudhuri:2014dla,  Dubovsky:2015cca, Aguilar-Arevalo:2016zop, Hardy:2016kme, Essig:2017kqs, 
Baryakhtar:2017ngi,Cardoso:2017kgn,
Baryakhtar:2018doz, Cardoso:2018tly, Pierce:2018xmy}.

In this work we will consider---from the low-energy effective theory
point of view---an even more minimal possibility than the dark photon,
namely, that the SM photon  is the new physics mediator.  
Beyond carrying a millicharge, DM may also interact \textit{directly} with
the photon through a number of higher dimensional operators that
encapsulate magnetic or electric dipole moment interactions (MDM or
EDM), an anapole moment (AM) or a charge radius interaction
(CR). These possibilities were originally considered in
\cite{Pospelov:2000bq,Sigurdson:2004zp,Ho:2012bg} with further studies
on the phenomenology found in
\cite{Schmidt:2012yg,Kopp:2014tsa,Ibarra:2015fqa,Sandick:2016zut,Kavanagh:2018xeh,
  Trickle:2019ovy}. Motivated by the intense efforts to search for
sub-GeV dark sector states~\cite{Essig:2013lka,Battaglieri:2017aum},
the topic of form-factor interactions was recently revisited in detail
by some of us~\cite{Chu:2018qrm}.

In~\cite{Chu:2018qrm} we focused on prospects of detecting
electromagnetic (EM) form factor interactions of a dark sector Dirac
particle $\chi$ with mass at or below the GeV-scale. The direct
production of pairs $\chi \bar \chi$ was constrained with data from
BaBar~\cite{Aubert:2001tu}, NA64~\cite{Banerjee:2017hhz} and
mQ~\cite{Prinz:1998ua} and future improvements in sensitivity were
derived for Belle-II~\cite{Abe:2010gxa}, LDMX~\cite{Akesson:2018vlm}
and BDX~\cite{Battaglieri:2016ggd}. The direct sensitivity was then
compared with indirect probes such as electroweak precision tests,
flavor physics constraints, as well as with results from LEP and
LHC. It was found that, owing to the higher dimensionality of the
operators, high energy probes provide superior sensitivity. The
conclusions are independent of the lifetime of $\chi$, as long as its 
stability is guaranteed while traversing terrestrial detectors.

In contrast, if $\chi$ is long-lived, additional constraints from
cosmology, astrophysics, and direct DM searches apply, and
in~\cite{Chu:2018qrm} we have considered the most important ones that
are crucial in the MeV-GeV mass bracket of $\chi$. However, once we
allow the $\chi$-mass to drop into the keV-region, additional
constraints from the production of $\chi\bar\chi$-pairs in stars
become important~\cite{Raffelt:1996wa}.  In this work we complement
our previous results derived in~\cite{Chu:2018qrm} with astrophysical
limits that apply once the dark state is stable on a macroscopic time
scale, without necessarily demanding that that sub-MeV $\chi$
particles make up the DM.
We derive the limits from stellar cooling that arise from red giant
(RG), horizontal branch (HB) stars, and the Sun, and revisit our
calculation of the supernova bound, taking into account
all major production channels.

\begin{figure*}[ht]
  \centering
\includegraphics[width=\textwidth]{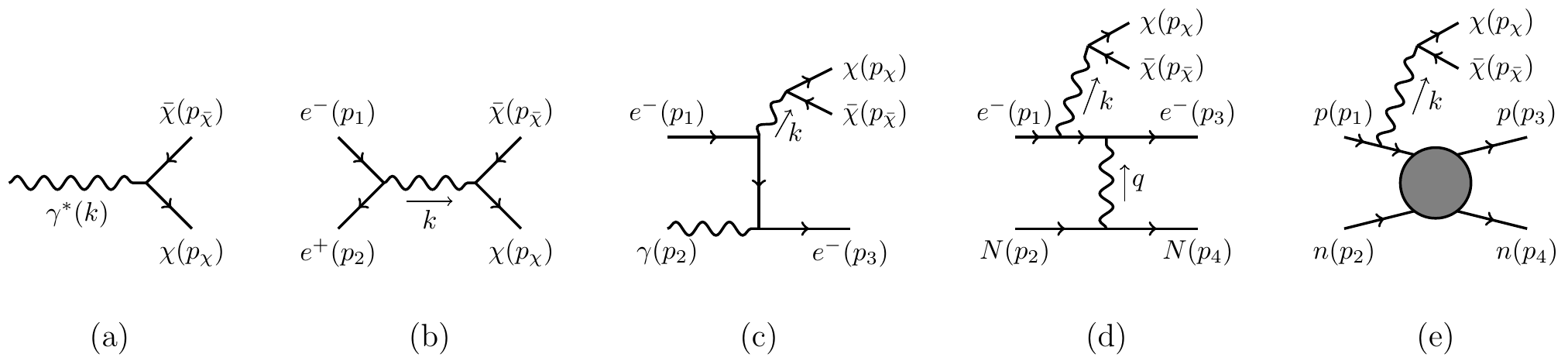}
\caption{Shown are the pair production processes of $\chi\bar\chi$
  that are calculated in this paper, namely, (a) plasmon decay, (b)
  $e^+e^-$ annihilation, (c) $2\to 3$ Compton scattering, (d) electron
  bremsstrahlung and (e) nucleon bremsstrahlung; for (c)-(d) we only
  show one of two relevant diagrams. The four momentum of the
  $\chi\bar\chi$-producing photon is denoted by $k$ throughout the
  paper.}
  \label{fig:feyn_diag}
\end{figure*}

Concretely, we are considering the following fundamental dark state
emission processes, highlighting in brackets the stellar system(s) for
which the process is most relevant,
\begin{align}
\label{eq:plasmon} & \!\!\!\text{Plasmon decay:} &&  \gamma_{\rm T,L}  \to  \chi  \bar\chi   &&  \text{(all)},   \\
\label{eq:annihilation} &\!\!\! \text{Annihilation:} && e^+  e^-  \to \chi  \bar\chi    &&   \text{(SN)},  \\
 & \!\!\!\text{Bremsstrahlung:}  && e^-  N  \to e^-  N   \chi   \bar \chi  &&   \text{(RG, HB, Sun)}, \!\!\! \nonumber \\
\label{eq:np} &   && NN  \to NN   \chi   \bar \chi  &&   \text{(SN)},\\
\label{eq:comp} & \!\!\!\text{Compton scattering:}  && e^-\gamma_{\rm T, L} \to e^- \chi  \bar \chi   &&   \text{(all)}. \!
\end{align}
The respective processes are decay of in-medium longitudinal (L) and
transverse (T) modes of thermal photons $ \gamma_{\rm T,L} $ which we
will simply refer to as ``plasmons'', electron-positron annihilation,
electron bremsstrahlung on protons and nuclei, nucleon-nucleon
bremsstrahlung and Compton scattering with the emission of a
$\chi\bar\chi$-pair. Exemplary respective diagrams are shown in
Fig.~\ref{fig:feyn_diag}.

The paper is organized as follows: in Sec.~\ref{sec:electr-form-fact}
we first set the stage by listing the effective operators that mediate
$\chi$-photon interactions. Section~\ref{sec:astr-observ} gives a
brief account on stellar energy loss arguments.  Our calculations on
$\chi$ particle emission are presented in
Sec.~\ref{sec:produc-DM}. The ensuing constraints are then collected
in Sec.~\ref{sec:constraint} before concluding in
Sec.~\ref{sec:conclusions}. Several appendices provide details on the
calculations and are referenced in the main text.

\section{Electromagnetic form factor interactions}
\label{sec:electr-form-fact}

A Dirac fermion $\chi$  may have a number of
interactions with the photon gauge field $A_{\mu}$ or its field
strength tensor~$F_{\mu\nu}$. At mass dimension-5 the interaction
terms of the Lagrangian are given by 
\begin{align}
  \label{eq:dim-5}
\mathcal{L}_{\chi}^{dim-5} = 
    \frac{1}{2} \mu_\chi \, \bar\chi \sigma^{\mu\nu} \chi F_{\mu\nu}   + \frac{i}{2} d_\chi \, \bar\chi  \sigma^{\mu\nu}\gamma^5 \chi F_{\mu\nu} ,
  \end{align}
  where $\mu_\chi$ and $d_\chi$ are the MDM and EDM coupling which may
  be measured in units of the Bohr magneton,
  $\mu_B \equiv e/(2m_e) = 1.93\times 10^{-11}\, e\,\cm $; $m_e$ is
  the mass of the electron and
  $\sigma^{\mu\nu} = \frac{i}{2} [\gamma^{\mu}, \gamma^{\nu}]$.  At
  mass dimension-6 we have  %
  \begin{subequations}
    \label{eq:dim-6}
  \begin{align}
    \mathcal{L}_{\chi}^{dim-6} & = - a_\chi \,  \bar\chi \gamma^{\mu} \gamma^5\chi \partial^{\nu} F_{\mu\nu}    + b_\chi \,  \bar\chi \gamma^{\mu} \chi \partial^{\nu} F_{\mu\nu} ,
\end{align}
\end{subequations}
where $a_{\chi}$ and $b_{\chi}$ are the AM and CR coefficients.  All
coupling strengths in Eqs.~\eqref{eq:dim-5} and \eqref{eq:dim-6} are real.
At mass-dimension-7 the interactions involve two photons at the vertex
and hence require a dedicated treatment. For this reason we restrict
our study to dim-5 and dim-6 operators.

The effective interactions in Eqs.~(\ref{eq:dim-5}) and (\ref{eq:dim-6})
may, \textit{e.g.},~arise from the compositness of
$\chi$~\cite{Bagnasco:1993st,Foadi:2008qv,Antipin:2015xia} or
perturbatively, from a UV completion that contains electrically
charged states~\cite{Raby:1987ga}. In the latter case, MDM and EDM
moments are \textit{e.g.}~generated by loop-induced axial or vector
Yukawa interactions $y_{A,V}$ of $\chi$ with additional scalars and
fermions. Parametrically, one expects
$\mu_{\chi} \sim Q |y_{A,V}|^2/M$ and
$d_{\chi}\sim Q \imag [y_V y_A^{*}]/M$ where $Q$ is the electric
charge of the mediator and $M$ is some common mass-scale of these new
states. In turn, the strength of AM and CR interactions may be expected as
$a_{\chi},\, b_{\chi} \sim Q |y_{A,V}|^2/ M^2$. It should be noted,
however, that these estimates may be significantly enhanced by the
lightness and/or mass-degeneracy of the spectrum of
states~\cite{Pospelov:2008qx}; a systematic study on EDMs induced by
CP violation from light dark sectors was recently performed
in~\cite{Okawa:2019arp}. In what follows, we treat the
interactions (\ref{eq:dim-5}) and (\ref{eq:dim-6}) independent of their
embedding.

For the Feynman-diagrammatic computation, one assembles the
interactions into the matrix element of the effective EM current
of~$\chi$, %
\begin{align*}
  \me{\chi(p_f)}{J_{\chi}^{\mu}(0)} {\chi(p_i)} & =  \bar u(p_f) \Gamma_{\chi}^{\mu}(q) u(p_i)  ,
\end{align*}
where $p_{i,f}$ and $q = p_i-p_f$ are four-momenta. For a neutral
particle $\chi$ %
the vertex functions reads,
\begin{align*} %
\begin{split}
  \Gamma_{\chi}^{\mu}(q) = & i \sigma^{\mu\nu} q_{\nu} \left(
    \mu_{\chi} + i d_{\chi} \gamma^5 \right) + \left( q^2 \gamma^{\mu}
    - q^{\mu}
    \slashed q \right) \left( b_{\chi} - a_{\chi}\gamma^5 \right). %
\end{split}
\end{align*}
Here we regard the various moments as being generated at an energy
scale well above the energies involved in the stellar production; they
are hence $q$-independent.

\section{Stellar observables}
\label{sec:astr-observ}

In this section we review the arguments on stellar energy loss.
Active stars such as RG, HB, or the Sun are systems of negative heat
capacity: if energy is lost, either through photon emission or through
new, anomalous processes, the decrease of total energy causes the
gravitational energy to become more negative. By virtue of the virial
theorem, the average kinetic energy and thereby the photon temperature
increases. The system heats up leading to a faster consumption of its
nuclear fuel while the overall stellar structure remains largely
unchanged.  In contrast, dead stars such as white dwarfs or the
proto-neutron star formed in core-collapse SN are supported by
degeneracy pressure and stellar energy loss implies a cooling of the
system. Constraints are then derived based on an observationally
inferred cooling curve.

\subsection{RG and HB stars}

In globular clusters (GCs), the population of stars on the red giant
branch vs.~horizontal branch is directly related to the lifetime of
stars in the respective phases. Their observationally inferred number
ratio agrees with standard predictions to within 10\%. Anomalous
energy losses shorten the helium-burning lifetime in HB stars,
creating an imbalance in the number of HB vs.~RG stars. This
constrains the luminosity in non-standard channels to be less than
approximately 10\% of the standard helium-burning luminosity of the HB
core~\cite{Raffelt:1996wa},
\begin{align}
  \label{eq:criterionHB}
  \int_{\rm core} dV\, \dot Q  <  10\% \times L_{\rm HB}  \quad \text{(HB)} \, .
\end{align}
Following \cite{Raffelt:1996wa},  $ L_{\rm HB}$ will be taken as $20\,L_\odot$ for a $0.5M_\odot$
core below. The values of the Solar mass and luminosity are
$M_\odot = 1.99\times 10^{33}\, \mathrm{g} $ and
$L_\odot = 3.83\times 10^{33}\, \mathrm{erg/s} $, respectively. The
computation of the anomalous energy loss rate per unit volume and time,
$\dot Q$, will be the subject of the next section.

A constraint for RG stars may be derived from an agreement between
predicted and observationally inferred core masses prior to helium
ignition. Energy loss delays the latter and the core mass keeps
increasing as the hydrogen burning ``ashes'' fall onto the degenerate
He core. Preventing an increase in core mass by no more than 5\%
yields the constraint~\cite{Raffelt:1996wa},
\begin{align}
  \label{eq:criterionRG}
   \dot Q  <  10 \,\mathrm{erg/g/s}  \times \rho \quad \text{(RG)} . 
\end{align}
Here, $\dot Q$ is to be evaluated at an average density of
$\rho = 2\times 10^5\, \mathrm{g}/\mathrm{cm}^3$ and a temperature of
$T=10^8\,\mathrm{K} \simeq 8.6\,\keV$, slightly higher than that of HB
stars. %

The criterion (\ref{eq:criterionRG}) on energy loss can be improved
utilizing high precision photometric observations of GCs. For example,
considering the brightness of the tip of the RG
branch,~\cite{Viaux:2013lha} has provided a detailed error budget and
new limits on neutrino dipole moments from GC M5 were derived based on
predictions of absolute brightness in the presence of anomalous energy
loss that are obtained with dedicated stellar evolutionary codes.  It
was found, however, that previously derived limits based on
(\ref{eq:criterionRG}) remain largely intact, as there appears to be
a slight preference for anomalous energy loss
channels~\cite{Viaux:2013lha}. In the following, for our purposes it
will hence be entirely sufficient to employ the simple
condition~(\ref{eq:criterionRG}) to arrive at constraints on the
EM form factors.

\subsection{Sun}

Solar neutrino fluxes are a direct measure of the nuclear fusion rates
inside the Sun.  For example, not only  the ${}^8$B neutrino flux is very well
measured but also the sensitive dependence of the responsible reaction on 
temperature provides an excellent handle for constraining anomalous
energy losses. The ensuing constraint is then phrased in terms of the
total Solar photon luminosity~\cite{Frieman:1987ui,Raffelt:1988rx}, as 
\begin{align}
  \label{eq:criterionSun}
 \int_{\rm Sun} dV\, \dot Q < 10\%  \times L_{\odot} \quad (\text{Sun}) .
\end{align}
It is important to note that (\ref{eq:criterionSun}) is basically
insensitive to the long-standing ``solar opacity problem'': the
measured ${}^8$B neutrino flux is situated in the overlap region of the nominal
error ranges between the discrepant high- and low-metallicity
determinations of the Solar chemical
composition~\cite{Redondo:2013lna}; see the respective references
\cite{Grevesse:1998bj} and \cite{Asplund:2009fu}. Hence,
(\ref{eq:criterionSun}) suffices as a criterion, awaiting further
developments on Solar opacity determinations.

\subsection{Supernova}

New particles that are emitted from the proto-neutron star and that
stream freely may quench the electroweak rates of neutrino emission
during the cooling phase. The involved processes and their dynamics
are highly complex. However, an approximate but very useful criterion
to constrain additional energy loss is the condition that the total
luminosity due to non-standard processes  should not exceed the neutrino
luminosity at one second after core bounce~\cite{Raffelt:1996wa},
\begin{align}
  \label{eq:criterionSN}
 \int_{\rm core} dV\, \dot Q < L_{\nu} = 3\times 10^{52}\, \mathrm{erg}/s \quad \text{(SN)}. 
\end{align}
The applicability of the bounds above are contingent on
that the SN1987A was a neutrino-driven SN
  explosion\footnote{For an alternative explosion mechanism where the
     SN1987A bounds would not apply, see \cite{Blum:2016afe,
      Bar:2019ifz}. } and that the produced particles are able to
escape the dense environment of the SN remnant, assumed to be
a  proto-neutron star (PNS).  %
Below, we will account for this so-called ``trapping-limit'' in the
case of SN. For all other systems introduced above, trapping is either irrelevant, or
happens in a parameter region that is excluded otherwise.

\section{Production cross sections and energy loss rates}
\label{sec:produc-DM}

In this section we first provide the general formula for $\chi\bar\chi$
pair production in the thermal bath, before breaking it down into the
most relevant pieces that dominate the in-medium production cross
sections and, thereby, the stellar cooling rates.

 \subsection{Exact formula for \texorpdfstring{\boldmath$\chi\bar\chi$}{chi} pair production}
\label{sec:general}

In thermal field theory, the production rate of a decoupled fermion
per volume per time may be obtained from its relation to the imaginary
part of its self-energy in medium~\cite{Weldon:1983jn} via
\begin{equation}
	\dot N_{\chi}= -\int \frac{d^3 \vec p_\chi}{(2\pi)^3} {1\over (e^{E_{\chi}/T}+1)}\,{\imag\Pi_{\chi} (E_{\chi}, \vec p_{\chi}) \over E_{\chi}} \,,
\end{equation}
where  $\imag\Pi_{\chi}(E_{\chi}, \vec p_{\chi}) = \bar u(p_{\chi})
\Sigma(E_{\chi}, \vec p_{\chi}) u(p_{\chi}) $ is the discontinuity of
the thermal self-energy of $\chi$, $\Sigma(E_{\chi}, \vec p_{\chi})$;
$u(p_{\chi})$ and $\bar u(p_{\chi})$ are free particle spinors with
four-momentum $p_{\chi} = (E_{\chi}, \vec p_{\chi})$.
To lowest order in the dark coupling,
$\Sigma(E_{\chi}, \vec p_{\chi})$ is found from the one-loop diagram
with a dressed photon propagator attached to the $\chi$ fermion line.
A general exposition on calculating discontinuities in the thermal
plasma is found in~\cite{Weldon:1983jn, Carrington:2002bv}.

Below, in Eq.~(\ref{eq:generalRate}), we are using a different
formulation and the equivalence may be appreciated in the following
way: when cutting the self-energy diagram for $\chi$, the optical
theorem implies that the production rate may also
be obtained by computing all graphs where a photon $\gamma^{*}$ of
four-momentum $k = p_{\chi}+ p_{\bar\chi}$ emerges from a SM current
and is being dotted into the dark current of the $\chi\bar\chi$
pair. The SM-process that leads to the creation of $\gamma^{*}$ is in
turn related to the imaginary part of the photon self-energy in the
medium, $\imag\Pi_{\mu\nu}$, where 
\begin{equation}
  \label{eq:PIdecomp}
  \Pi^{\mu\rho} =  (  \epsilon_{\mathrm{T},1}^{\mu}  \epsilon_{\mathrm{T},1}^{\rho}
  + \epsilon_{\mathrm{T},2}^{\mu}  \epsilon_{\mathrm{T},2}^{\rho})   \,\Pi_\mathrm{T} +  \epsilon_\mathrm{L}^{\mu}  \epsilon_\mathrm{L}^{\rho} \,\Pi_\mathrm{L}\,.
\end{equation}
Here $\epsilon_{\mathrm{T,L}}$ are the transverse and longitudinal
photon polarization vectors and $\Pi_{\mathrm{L},\mathrm{T}}(k) $ is
thermal photon self-energy for the respective polarization; explicit
expressions are given in App.~\ref{sec:phot-therm-medi}. Identifying
the leading contributions to $\imag\Pi_{\mathrm{L},\mathrm{T}}(k) $ in various mediums  
then allows to account for the dominant $\chi$ pair production
channels.

The exact differential production rate per volume of $\chi\bar\chi$ pairs via a
photon of 4-momentum $k = (\omega, \vec k)$ emerging from \textit{any}
SM process to lowest order in the dark current can be obtained by
borrowing the results from dilepton production in hot matter,
see~\textit{e.g.}~\cite{Alam:1999sc,Bellac:2011kqa}. Adopted to our
purposes (see App.~\ref{sec:appdecay-rate-cross}) it reads,
\begin{eqnarray}\label{eq:generalRate}
	 { d\dot{N}_{\chi} \over d\sdark}  &=& -
\sum_{i= {\rm T, L}} g_i \int \frac{d^3 \vec k}{(2\pi)^3} { 1 \over  (e^{\omega/T}-1) } {\imag\Pi_{i}(\omega, \vec k)\over \omega} \notag \\
  &&\times    {f(\sdark)  \over 16\pi^2 |\sdark-\Pi_{i}|^2}   \sqrt{1-\frac{4 m_\chi^2}{\sdark}} \, ,
\end{eqnarray}
where $\sdark=k^2$ is the invariant mass of the $\chi$-pair and the
internal degrees of freedom of two polarization modes are
$g_{\rm T} =2, \,g_{\rm L}=1$.
The differences in the various interaction possibilities are entirely
captured in a factor that will repeatedly appear and that was obtained
in our preceding work~\cite{Chu:2018qrm},
\begin{subequations}\label{eqn:f}
\begin{align} 
\label{eqn:f_MDM}
	\text{MDM:\quad } 
	f(\sdark) &= \frac{2}{3} \mu_\chi^2 \sdark^2 \left( 1+\frac{8 m_\chi^2}{\sdark}\right) , \\
\label{eqn:f_EDM}
	\text{EDM:\quad } 
	f(\sdark) &= \frac{2}{3}d_\chi^2 \sdark^2 \left( 1-\frac{4 m_\chi^2}{\sdark}\right) , \\
\label{eqn:f_AM}
	\text{AM:\quad } 
	f(\sdark) &=  \frac{4}{3} a_\chi^2  \sdark^3 \left(1-\frac{4 m_\chi^2}{\sdark}\right) ,\\
\label{eqn:f_CR}
	\text{CR:\quad } 
	f(\sdark) &=  \frac{4}{3} b_\chi^2  \sdark^3 \left(1+\frac{2 m_\chi^2}{\sdark}\right) .
\end{align}
\end{subequations}
Equation \eqref{eq:generalRate} is the general expression of the
weakly coupled $\chi$ pair-production rate from the thermal
medium;  details are found in App.~\ref{sec:appdecay-rate-cross}.

The contribution to $\chi\bar\chi$ production to leading order in
$\alpha$ is given by the \emph{pole} in~\eqref{eq:generalRate},
\textit{i.e.}~for $\sdark = \real\Pi_{\mathrm{L},\mathrm{T}}$.
When this condition is met,~\eqref{eq:generalRate} reduces to the
decay rate of thermal photons
$\gamma_{\mathrm{L},\mathrm{T}}\to \chi\bar\chi$. Hence, resonant
$\chi\bar \chi$ production is fully accounted for by
$\gamma_{\mathrm{L},\mathrm{T}}$ decay. The decay itself becomes
possible by virtue of the in-medium (squared) mass of
$\gamma_{\mathrm{L},\mathrm{T}}$: it is given by
$\real\Pi_{\mathrm{L},\mathrm{T}}(\omega_{\rm L, T}, \vec k)$, where
$\omega_{\rm L,T}$ %
denotes the solution of $\omega(|\vec k|)$ of the corresponding
longitudinal and transverse dispersion relations
$\omega^2 -{|\vec k|}^2 -\real\Pi_{\mathrm{L},\mathrm{T}}(\omega, \vec
k)=0$. Plasmon decay is discussed in the following subsection, and
explicitly calculated in
(\ref{eq:appgeneralRate}-\ref{eq:decaygeneral}) in
App.~\ref{sec:appdecay-rate-cross}.  %
The expressions for $\real\Pi_{\mathrm{L},\mathrm{T}} $ and
finite-temperature dispersion relations are found in
(\ref{Eq:Pol_tensor}-\ref{Eq:dispersion_relation}).

\begin{figure*}[t]
  \centering
\includegraphics[width=\textwidth]{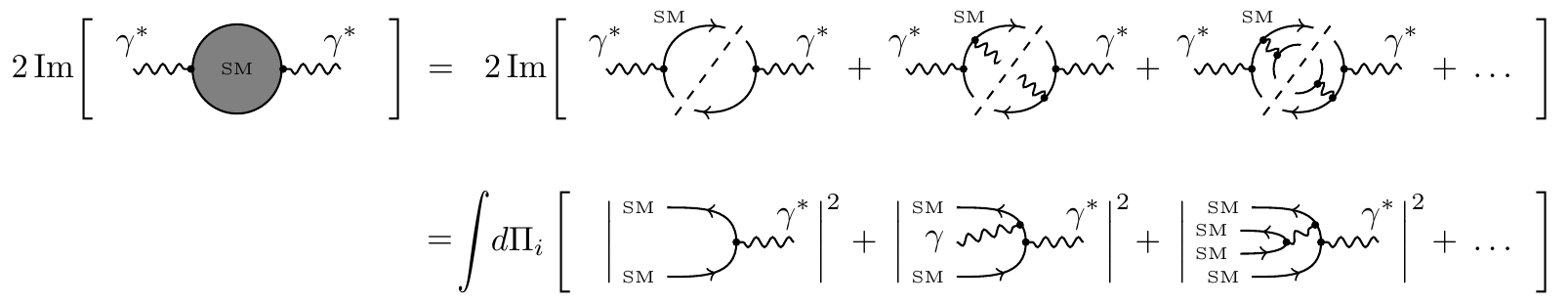}
\caption{Optical theorem relating the imaginary part of the photon
  self energy to the sum of all SM processes that create an off-shell
  photon $\gamma^{*}$. The first equality shows the leading individual
  contributions to the self-energy. When the latter loop-diagrams are
  cut, they correspond to the scattering processes shown in the second
  line, where $d\Pi_i$ symbolizes the phase space integral of all
  external particles, except $\gamma^*$. When the scattering diagrams
  are deformed in a way such that two SM particles are in the initial
  state, the processes correspond to annihilation, Compton scattering
  and bremsstrahlung (from left to right). Any diagrams with $\chi$
  particles involved yield contributions to the production rate
  (\ref{eq:generalRate}) that are of higher order in the dark
  coupling.}
  \label{fig:optical}
\end{figure*}

Production \emph{off-the-pole} to Eq.~\eqref{eq:generalRate} can be
elucidated by studying the contributions to $\imag{\Pi}$ using the
optical theorem, illustrated in Fig.~\ref{fig:optical}. The left hand
side shows the fully dressed vacuum polarization of an off-shell
photon $\gamma^{*}$, found by considering loop-diagrams of increasing
order in $\alpha$ illustrated in the first equality. When those loop
diagrams are cut, their imaginary parts are given by the tree-level
production processes for $\gamma^*$ shown in the last equality.
The leading $\alpha$ contribution to $\imag{\Pi}$ is then given by the
electron one-loop diagram. Although it is well known that on-shell
plasmon decay $\gamma_{\mathrm{L},\mathrm{T}} \to e^+e^-$ remains forbidden at finite
temperature~\cite{Braaten:1990de}, an electron loop still contributes
to $\imag{\Pi}$ in the off-shell case. The associated process is then
$e^+e^-$ annihilation to $\chi\bar\chi$, i.e.~process (\ref{eq:annihilation}b).

The second and third diagrams in the last
line of Fig.~\ref{fig:optical} are related to $\chi\bar\chi$
production in Compton scattering and bremsstrahlung. Here, it is
important to note that $\sdark = \real\Pi_{\mathrm{L},\mathrm{T}}$ can
also be met in the photon propagator that produces the $\chi$-pair
with invariant squared mass $\sdark$. However, including such
resonances would amount to double-counting. As we have seen above, the
pole contributions are already captured by plasmon decay.%
\footnote{A heuristic argument on such double counting was also given
  in the context of neutrino pair emission in Sec.~2.5 of
  \cite{Vitagliano:2017odj}. It furthermore appears to us, that double
  counting may have occurred in \cite{Chang:2018rso} where a
  potentially resonant bremsstrahlung process was added to the plasmon
  decay contribution.}
In our calculations, we explicitly avoid this situation by setting
$\Pi_{\mathrm{L},\mathrm{T}}\to 0$ in the propagator if the resonance is kinematically
allowed for the photon that directly couples to the dark current. We
have numerically verified that our results remain otherwise unaffected
by neglecting the thermal shift in the photon propagator.

Finally, there is also a potential double counting between Compton
scattering and bremsstrahlung processes, which happens when in the
bremsstrahlung process the photon exchanged between two initial
particles carries 4-momentum $q$ (see Fig.~\ref{fig:feyn_diag}d) that
satisfies the dispersion relation $q^2-\real\Pi_{L}(q^0,\vec q)=0$,
leading to the  exchange of an on-shell longitudinal plasmon.
The process then becomes equivalent to Compton scattering
$e/N + \gamma_{\mathrm{L}} \to e/N + \chi + \bar \chi $. This has been
reported for axion production processes, where the contribution of
latter is mostly covered by that of
bremsstrahlung~\cite{Raffelt:1987np}.  To avoid such double-counting,
we take the static approximation ($q^0=0$) for the thermal mass of the
photon exchanged in bremsstrahlung processes, which is a valid limit
as the nucleon mass is large.  As $q^2<0$ and
$\Pi_{L} (q^0=0, \vec q)$ is always positive, the exchanged photon can
not become on-shell in bremsstrahlung processes, thus double counting
is avoided (see Sec.~\ref{sec:compton-scattering} for more details).

 \subsection{\texorpdfstring{\boldmath$\gamma_{\mathrm{T},\mathrm{L}}$}{gamma} decay}
\label{sec:plasmon-decay-rate}

The on-shell process of photon decay to $\chi\bar\chi$ (Fig.~\ref{fig:feyn_diag}a) %
 becomes possible in the medium and has an important analogy in
the literature, the plasmon decay to neutrinos.  Since the dispersion
relation for transverse  and longitudinal thermal photons are distinct, it is again
helpful to separate the two polarizations in the calculation. Explicitly, we
obtain for the decay rate per  degree of freedom
\begin{equation}\label{eq:decayT}
\begin{split}
&\Gamma_\mathrm{T, L} = \dfrac{1}{16\pi} Z_\mathrm{T, L } \sqrt{1-\dfrac{4 m_\chi^2}{\omega_{\rm T,L}^2-|\vec k|^2}} \dfrac{f(\omega_{\rm T,L}^2-|\vec k|^2)}{\omega_{\rm T,L}}\,,
\end{split}
\end{equation}
where $\omega_{{\rm T,L}} =\omega_{\rm T,L}(|\vec k|)$  for each polarization mode, as defined above. %
Details on the definition of the wave function renormalization factors
$Z_\mathrm{T, L }$ and the calculation are again given in
App.~\ref{sec:appdecay-rate-cross}. In the limit of $m_{\chi}\to 0$,
the decay widths for MDM agree with the well-known formul\ae\ for
plasmon decay to a neutrino pair~\cite{Raffelt:1996wa}.

For the plasmon decay processes, the energy loss
rate can be expressed as~\cite{Raffelt:1996wa},
\begin{equation}
\begin{split}
&\dot{Q}_{\mathrm{decay,T}} = \dfrac{2}{2\pi^2} \int^\infty_0 d|\vec k|\,  \dfrac{|\vec k|^2 \Gamma_\mathrm{T}\omega_\mathrm{T}}{e^{\omega_\mathrm{T} /T}-1} \Theta(\omega_{\mathrm{T}}^2 -|\vec k|^2 -4m_\chi^2),\\
&
\dot{Q}_{\mathrm{decay,L}} = \dfrac{1}{2\pi^2} \int^{k_\mathrm{max}}_0
\!\!\!\! d|\vec k|\, \dfrac{|\vec k|^2 \Gamma_\mathrm{L}\omega_\mathrm{L}}{e^{\omega_\mathrm{L} /T}-1} \Theta(\omega_{\mathrm{L}}^2 -|\vec k|^2 -4m_\chi^2).
\label{eq:Qplasmon}
\end{split}
\end{equation}
The expression for $k_{max}$ is given in Eq.~\eqref{Eq:kmax}.  For a
non-relativistic medium (HB, RG, Sun), the dispersion relation crosses
the light-cone at $|\vec k|=k_{max}$, signaling the damping of
longitudinal modes ($i.e.$~Landau damping); for a relativistic plasma (SN)
$k_{max} \to \infty$.  The relative factor of $2$ between the
expressions reflects the counting of polarization degrees of freedom.
Finally, the last factor is a kinematic restriction on the phase
space, $\omega_{\mathrm{T},\mathrm{L}}^2 - |\vec k|^2 \ge 4m_\chi^2$.
For transverse mode thermal photons, the integral becomes bounded from below since
$\omega_\mathrm{T}^2 -|\vec k|^2$ increases as $|\vec k|$ increases
according to the dispersion relation.
For the longitudinal case, the integral is additionally
bounded from above since the trend in
$\omega_\mathrm{L}^2 -|\vec k|^2$ with respect to $|\vec k|$ is
reversed.

\subsection{\texorpdfstring{\boldmath$e^+ e^-$}{ee} annihilation}
\label{sec:epem-anni-prod}
The degenerate plasma of the PNS core with temperature
$T\gg m_e$ contains a population of $e^+$, allowing for dark state
pair-production through $e^+e^-$ annihilation
(Fig.~\ref{fig:feyn_diag}b). %
The calculation for the pair production cross section is detailed in
App.~\ref{sec:appdecay-rate-cross}.

In terms of the invariant $s=(p_1+p_2)^2$ and the sum/difference 
  of incoming $e^\mp$ energies $E_{1,2}$ in the frame of the thermal
bath, \textit{i.e.}, $E_\pm \equiv E_1 \pm E_2$, the corresponding
cross section mediated by the transverse polarization part of the
propagator reads,
\begin{equation}
\label{Eq:Final_ann_cross_T}
 \sigma_\mathrm{T} = \dfrac{\alpha\left[sE_-^2+(4m_e^2+s)(E_+^2 -s)\right]}{8 \sqrt{s(s-4m_e^2)}(E_+^2 -s)(s-\Pi_T)^2} \sqrt{1-\dfrac{4m_\chi^2}{s}}f(s). 
\end{equation}
For the longitudinal part we obtain
\begin{equation}
\label{Eq:Final_ann_cross_L}
\sigma_\mathrm{L} = \dfrac{\alpha \left[ s(E_+^2  -E_-^2-s)\right]}{8 \sqrt{s(s-4m_e^2)}(E_+^2 -s)(s-\Pi_L)^2} \sqrt{1-\dfrac{4m_\chi^2}{s}}f(s).
\end{equation}
Note that there is no interference term between the two. Furthermore,
the sum of both cross sections,
$ \sigma_\mathrm{T} + \sigma_\mathrm{L}$, becomes Lorentz invariant in
the limit of $\Pi_{\mathrm{T},\mathrm{L}} \to 0$.%
\footnote{ We use a definition of the cross section for which the
  M{\o}ller velocity instead of the relative velocity
  $|\vec v_1 - \vec v_2 |$ appears. At zero temperature, this makes
  the cross section a Lorentz invariant quantity, see the discussion
  in~\cite{Gondolo:1990dk}.}

Before using (\ref{Eq:Final_ann_cross_T}) and
(\ref{Eq:Final_ann_cross_L}) in the calculation of the energy loss
rate, a comment on the analytic structure is in order. Although it
appears that the process may be significantly enhanced when
$s = \real \Pi_{\mathrm{T},\mathrm{L}}$, this condition is never met:
for the same reason that the decay of thermal photons into an
electron-positron pair ($\gamma_{\mathrm{T},\mathrm{L}}\to e^+e^-$) is
forbidden~\cite{Braaten:1990de}, the finite-temperature corrections to
$m_e$ prevent the process (\ref{eq:annihilation}) from going
on-shell. It is for this reason that we have explicitly evaluated the
thermal electron mass for the employed radial profile of the
PNS; see App.~\ref{sec:phot-therm-medi}. In other
words, we use a thermal electron mass in SN, and use the
zero-temperature electron mass in HB, RG and Sun, where $e^+e^-$
annihilation is of little relevance. 
The values of chemical potential $\mu_e$ are self-consistently
adjusted to match the numerical PNS profiles from the literature (see below).

The energy loss rate of $e^+ e^-$ annihilation is found by weighing
the emission process by the total radiated final state energy
$E_3 + E_4 = E_1 + E_2$ and by the probability of
finding the initial states with the respective energies $E_1$ and $E_2$,
\begin{equation}
  \begin{split}
    \label{eq:Qann}
    \dot{Q}_{\mathrm{ann}} =& \int d\Pi_{i=1,2,3,4} (2\pi)^4 \delta^4(p_1 + p_2 -p_3 - p_4)
    \\ &\times g_{e^-}g_{e^+}f_{e^-} f_{e^+} \dfrac{1}{4} \sum\limits_{\rm spins} |\mathcal{M}_\mathrm{ann}|^2 (E_1 +E_2).
\end{split}
\end{equation}
Here, $f_{e^\pm}$ are the phase-space distributions of $e^\pm$, with
internal degrees of freedom $g_{e^\pm} =2$, and
$|\mathcal{M}_\mathrm{ann}|^2$ is the squared matrix element for
$e^+ e^-$ annihilation into the dark state
pair. In~(\ref{eq:Qann}) a Pauli-blocking factor
  induced by $\chi$ and $\bar\chi$ is neglected; we have verified that this does not
  affect the derived constraints.
Finally, $d\Pi_i = \prod_i d^3\vec p_i(2\pi)^{-3}(2E_i)^{-1}$ is the
Lorentz invariant phase space element.
The energy loss rate can be written in terms of the cross sections
$\sigma_{\rm T,L}$. Borrowing from the discussion on phase space
in~\cite{Edsjo:1997bg}, we find explicitly
\begin{equation}
\begin{split}
\dot{Q}_\mathrm{ann} &=  \int^\infty_{4 m_\mathrm{th}^2} \!\! ds \int^\infty_{\sqrt{s}} \! dE_+ 
\int^{\sqrt{(1-4 m_e^2 /s)(E_+^2 -s)}}_{-\sqrt{(1-4 m_e^2 /s)(E_+^2 -s)}} \!\!dE_- \\
&\times \dfrac{1}{64 \pi^4}g_{e^-}g_{e^+}f_{e^-} f_{e^+}  E_+ \sqrt{s(s-4 m_e^2)}  \sigma_{\mathrm{T},\mathrm{L}} \, .
\end{split}
\end{equation}
The distribution functions  $f_{e^-}$ and $f_{e^+}$ read
\begin{equation}
f_{e^\mp}  = \dfrac{1}{e^{(E_+ \pm E_- \mp 2\mu_e)/2T}+1}.
\end{equation}
Here, $\mu_e$ is the chemical potential of electrons and $T$ is the
temperature. 
The threshold mass $m_\mathrm{th}$ is equal to
$\mathrm{max} \lbrace m_e , m_\chi \rbrace$.

\begin{figure*}[tb]
\centering
\includegraphics[width=0.97\columnwidth]{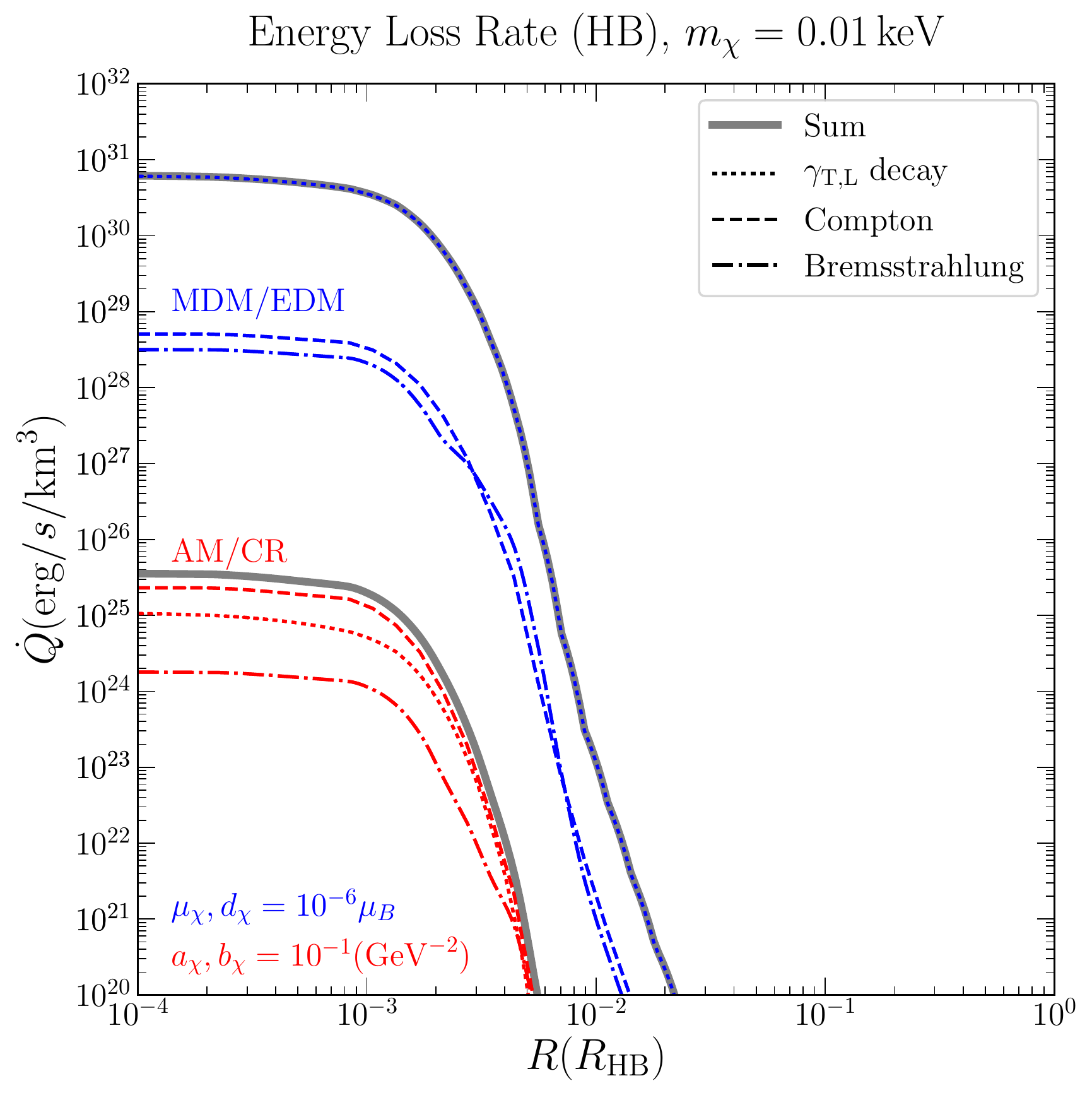}\hfill
\includegraphics[width=0.97\columnwidth]{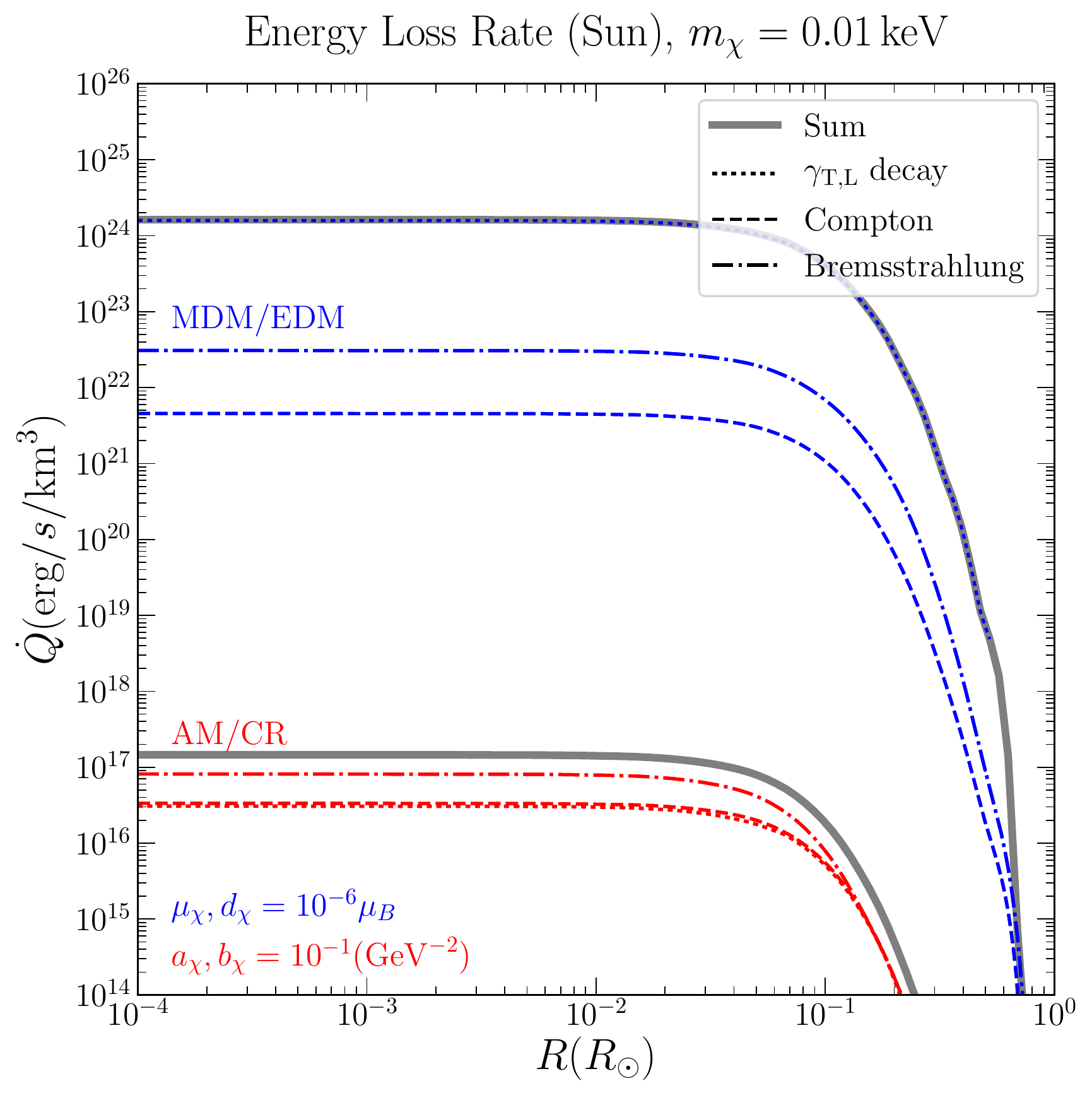}
\caption{\textit{Left:} Energy loss rates as a function of fractional
  stellar radius from $\gamma_{\rm T,L}$ decay (dotted lines), Compton
  production (dashed lines) and electron bremsstrahlung (dash-dotted lines) for
  $m_\chi = 0.01 \,\mathrm{keV}$ and
  $\mu_{\chi} (\text{or }d_{\chi}) = 10^{-6}\,\mu_B$ and
  $a_{\chi} (\text{or }b_{\chi}) = 0.1/\GeV^2$ in the representative HB star we
  consider. The sum of all processes is shown by the thick gray line,
  which for MDM/EDM interactions practically coincides with plasmon
  decay.  \textit{Right:} The same processes as in the left panel but
  for the Sun.}
\label{fig:ELR_HBsun}
\end{figure*}

\begin{figure*}[tb]
\centering
\includegraphics[width=\columnwidth]{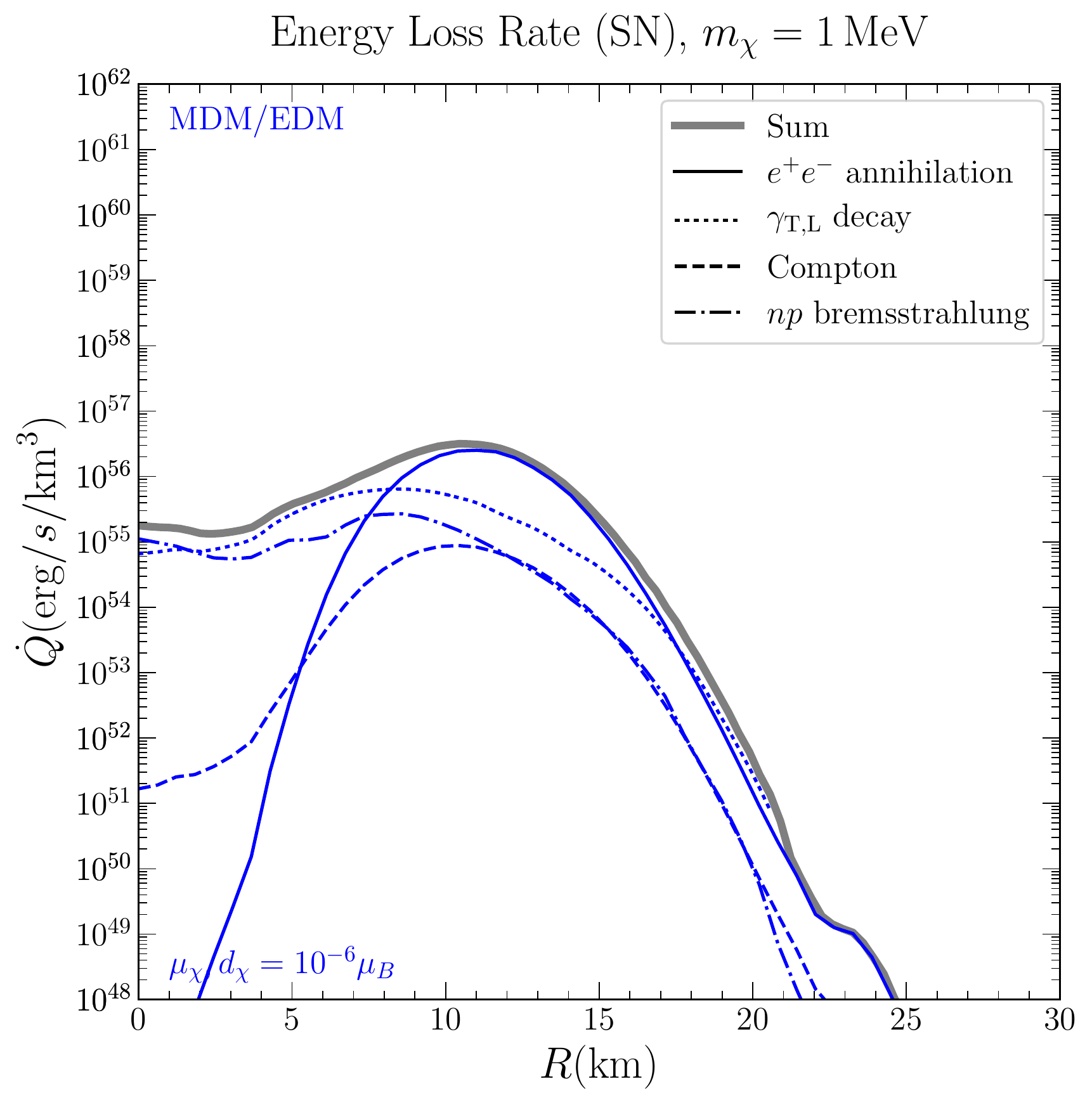}
\includegraphics[width=\columnwidth]{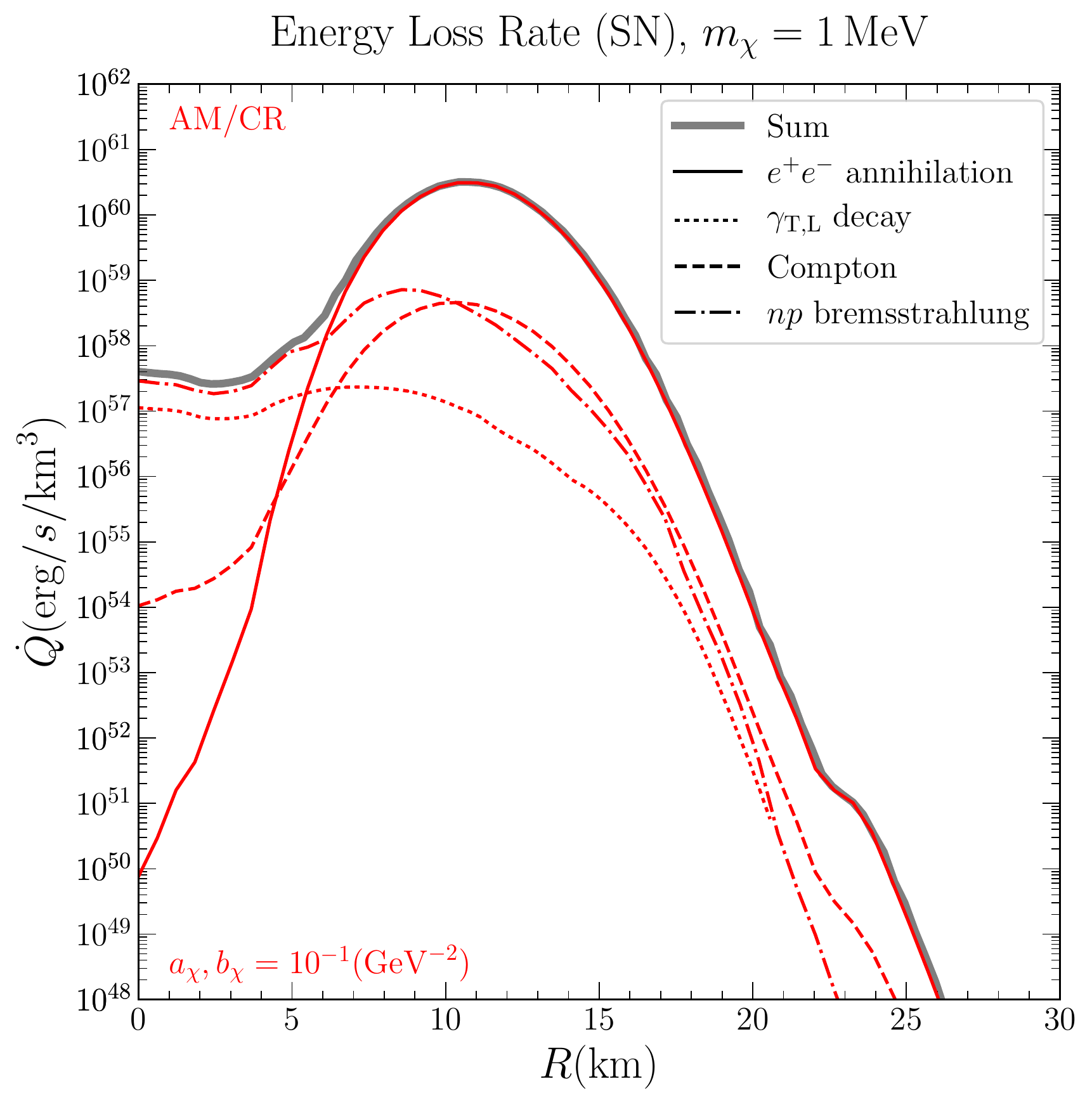}
\caption{{\it Left:} Energy loss rates inside PNS for MDM/EDM
  interactions with
  $\mu_{\chi} (\text{or } d_{\chi}) = 10^{-6}\,\mu_B$ and
  $m_\chi = 1 \,\mathrm{MeV}$ are shown for all computed processes,
  namely, $e^+ e^-$ annihilation (thin solid line), $\gamma_{\rm T,L}$
  photon decay (dotted line), Compton production (dashed line) and
  $np$ bremsstrahlung (dash-dotted line). The sum of all contributions is
  the thick solid line.  {\it Right:} The same processes as in the
  left panel but for AM/CR interactions with
  $a_{\chi} (\text{or } b_{\chi}) = 0.1/\GeV^2$. }
\label{fig:ELR_SN}
\end{figure*}

\subsection{Compton scattering}
\label{sec:compton-scattering}

For $2\to 3$ Compton scattering
($e^- /N + \gamma_{\rm T,L} \to e^- /N + \chi + \bar \chi $) with an
initial $\gamma_{\rm T,L}$ (Fig.~\ref{fig:feyn_diag}c), we calculate
the differential cross section via
\begin{equation}
	\dfrac{d\sigma_{2 \to 3}}{d \sdark} = \sigma_{2 \to 2}(\sdark) \dfrac{f(\sdark)}{16 \pi^2 \sdark^2} \sqrt{1- \dfrac{4m_\chi^2}{\sdark}}\,.
\end{equation}
Here, $\sigma_{2 \to 2}(\sdark)$ is the cross section of the two-body
Compton scattering with the final-state photon having a mass
$\sqrt{\sdark}$. We are only required to consider the process on
electrons, $e^- + \gamma_{\rm T,L} \to e^- + \chi + \bar \chi $, as
Compton scattering on protons is strongly suppressed.
Following the treatment in \cite{Vitagliano:2017odj} and our
discussion above, we neglect the thermal mass of the final state
photon in $\sigma_{2 \to 2}$ to avoid any potential double counting
with $\gamma_{\rm T, L}$ decay. For the initial state photon in the
integration of energy loss rate, Eq.~(\ref{eq:Qcompton}) below, the
thermal mass is properly taken in account through the dispersion
relation~\eqref{Eq:dispersion_relation}.

Furthermore, note that there is no double counting between the Compton
process and bremsstrahlung either in our treatment. A double counting would
appear if the $t$-channel photon exchange in bremsstrahlung, with 4-momentum $q$ (see Fig.~\ref{fig:feyn_diag}d), goes on
resonance. This is in principle possible for the longitudinal mode,
since $\Pi_{\rm L}$ in the propagator could become negative once the
dispersion relation of $\gamma_{\rm L}$ crosses the light cone.
Nevertheless, in the electron bremsstrahlung process discussed
below---most relevant for RG, HB and the Sun---the proton recoil and
hence the energy exchange are extremely small. Therefore, the
propagator can be taken in the static limit (energy exchange
$q^0 \rightarrow 0$).  This limit amounts to Debye screening, characterized by  
$\Pi_{\rm L} (q^0 \rightarrow 0, |\vec{q}|)$. Since
the screening scale is always positive, a resonance is never
met.  Therefore, we include the contribution from
$e^- + \gamma_{\rm L} \rightarrow e^- + \chi +\bar{\chi}$ to capture
the $t$-channel resonance contribution of electron bremsstrahlung, although it
is less important than plasmon decay. 

The energy loss rate from Compton scattering is calculated in a
similar way as~\eqref{eq:Qann}, but here $E_\text{loss}$ is given by
the energy carried by the virtual photon in the medium frame,
\begin{equation}
  \label{eq:Qcompton}
	\begin{split}
\dot{Q}_{\rm Compton} = &\int \!\! d\Pi_{i=1,2}   \, 4  E_1 E_2\,  \sigma^{\rm T,L}_{2\rightarrow 3} v_M\, g_{e^-} g_{\rm T,L} f_1  f_2  \\
  & \,\,\,\, \times(1-f_3)E_{\rm loss},
   \end{split}
\end{equation}
where $f_{1,2,3}$ are the distribution functions of the incoming
electron, $\gamma_{\rm T,L}$ and outgoing electron, respectively, with
$g_{e^-} = g_{{\rm T}}= 2$ and $g_{{\rm L}}= 1$ the internal degrees
of freedom for the incoming electron and $\gamma_{\rm T,L}$.  Pauli
blocking is accounted for by including the factor $(1- f_3)$.  The
energy loss $E_{loss} = E_{\bar{\chi}} +E_\chi$ can be expressed in
terms of variables defined in the medium frame.  Moreover, for RG, HB
and the Sun, the relativistic corrections induced by transforming from the 
center-of-mass (CM) frame to the medium frame are very small, and are neglected for
simplicity.

\subsection{\texorpdfstring{\boldmath$e^- N$}{eN} bremsstrahlung}
\label{sec:bram-dm}

In this subsection we consider dark state pair production from
bremsstrahlung of electrons on protons or other nuclei
(Fig.~\ref{fig:feyn_diag}d). %
Similar to the Compton scattering above, we also relate the
$2\rightarrow 4$ cross section to a $2\to 3$ process of
$eN\rightarrow eN+\gamma^*_{T,L}$ in which the emitted photon,
$\gamma^*_{T,L}$, has an invariant mass
$\sqrt{\sdark}$ %
\begin{equation}
\label{Eq:2to4to3}
\dfrac{d\sigma_{2\rightarrow 4}}{d\sdark} =
\sigma_{2\rightarrow 3} (\sdark)
\dfrac{f(\sdark)}{16\pi^2 \sdark^2} \sqrt{1-\dfrac{4m_\chi^2}{\sdark}} .
\end{equation}
In the following we shall only consider photon-emission from
electrons, as the emission from the nuclear leg is suppressed by a
factor of $(Z m_e/ m_N)^2 \ll 1$ where $Z$ and $m_N$ are the charge
and mass of the nucleon/nucleus. Furthermore, ordinary electron-electron
bremsstrahlung is a quadrupole emission process and correspondingly
smaller in practice. We therefore also neglect such production channel.

The $eN$ process is sensitive to the details of in-medium corrections.
To this end, recall that the $t$-channel photon exchange in 
Fig.~\ref{fig:feyn_diag}d has a well-known Coulomb divergence in the
limit of vanishing momentum-transfer. This issue is mitigated by two
factors: first, the divergence is not met kinematically as long as
$m_{\chi}\neq 0$ since a minimum momentum transfer is necessary
to create the final state pair. Second, the medium itself regulates
the process through the Debye screening of bare charges characterized by a
momentum scale $k_D$. The latter appears as the static limit
of $\Pi_\mathrm{L}(q^0\to 0, \vec q)$ and for a classical plasma reads,
\begin{equation}
\label{Eq:screening_scale}
k_D^2 = \dfrac{4\pi \alpha n_e}{T} + \text{ion-contributions}.
\end{equation}

For the numerical results, we have calculated
$\sigma_{2 \rightarrow 3} $ in (\ref{Eq:2to4to3}) using the 
propagator (\ref{Eq:Propagator_Coulomb}), neglecting, for simplicity,
ion contributions. We separate the squared amplitude into
transverse and longitudinal parts and include the static limits of
$\Pi_{\rm T,L}$ in the respective propagators.  For the longitudinal
part, the zero-temperature propagator $q^{-2}$ is replaced by
$(q^2 - k_D^2)^{-1}$. In contrast, there is no magnetic screening in
the static limit ($\Pi_\mathrm{T}(q^0 \rightarrow 0, |\vec q|) = 0 $),
hence there is no thermal screening for the propagator of the
transverse mode.
We find that in the non-relativistic limit the contribution of the
longitudinal mode dominates.

To avoid any double counting between this process and
$\gamma_{\rm T,L}$ decay, we need to subtract the contribution when
the virtual photon that directly couples to $\chi$ goes on-shell. As
stated above, this is achieved by setting $\Pi_{\rm T,L}$ in the
corresponding propagator to zero. Since this should over-estimate the
production rate at $\sdark \le \Pi_{\rm T,L}$, we have also tested an
opposite option of choosing
$\Pi_{\rm T,L} \to - \Pi_{\rm T,L} (E_\chi + E_{\bar \chi})$ to avoid
the singularity, which under-estimates the production rate. We find
that both prescriptions lead to same results at percent level, which
justifies our simplification of taking $\Pi_{\rm T,L}\equiv 0$ for the
photon that directly couples to $\chi$.

For dark state pair production in  $e^-$ bremsstrahlung on protons
and nuclei, the energy loss rate is expressed as
\begin{equation}
\begin{split}\label{eq:BremQ}
\dot{Q}_{\rm brem} = &\int \!\! d\Pi_{i=1,2}   \, 4  E_1 E_2\,  \sigma_{2\rightarrow 4} v_M\, g_1 g_2 f_1 f_2  (1-f_3)E_{\rm loss},
\end{split}
\end{equation}
where $f_{1,2,3}$ are the distribution functions of the incoming
electron, proton/nucleus and outgoing electron, with $ g_{1,2}$ the
internal degrees of freedom for the incoming particles.
We have neglected the Pauli blocking factor for final-state protons/nuclei as it plays little
role.
The M{\o}ller velocity $v_M= F/(E_1E_2)$ is given in terms of the flux
factor $F$ found in~\eqref{Eq:flux_factor}.
The energy carried-away by the dark states is
$E_{\rm loss}=E_{\bar{\chi}} + E_\chi$ and its expression 
in the medium frame is introduced in
App.~\ref{sec:appdecay-rate-cross}.

Making the approximation that protons and other nuclei are at rest,
their phase-space integral gives $\int d\Pi_2 f_2 = n_N/(2 m_N g_2)$,
where $n_N$ is the number density of the protons/nuclei.
Hence we arrive at
\begin{equation}
\begin{split}
\dot{Q}_{\rm brem} = \!\int_{m_e + 2 m_\chi}^\infty \!\!\!\!\!\!\!\!\!\!\!\!\!\! dE_1 \dfrac{2 n_N E_1 E_2 v_M}{(2\pi)^2 m_N} |\vec{p}_1| g_1 f_1  \sigma_{2\rightarrow 4} (1-f_3) E_{\rm loss},
\end{split}
\end{equation}
with $|\vec{p}_1| = \sqrt{E_1^2 -m_e^2}$ and where
$\sigma_{2\rightarrow 4}$ is obtained from integrating
(\ref{Eq:2to4to3}) over appropriate boundaries (see
App.~\ref{sec:bremsstr-prod-dark}). 
Generically, bremsstrahlung is less effective when pair annihilation or
plasmon decay are open as production channels, but it can be dominant at low temperatures where the latter processes are kinematically suppressed.

Before ending this subsection, it is worth commenting on the so-called
soft photon approximation, which states that in the limit that the emitted
photon energy is small compared to the available kinetic
energy (i.e., $\omega \ll E_\text{kin}$), the
 process of $eN\rightarrow eN+\gamma^*_{T,L}$  factorizes into a
product of elastic scattering times a factor describing the additional
emission of $\gamma^*_{T,L}$.
While this approximation works well for the emission of a massless
photon, it breaks down if the off-shell photon's {\emph{effective}} mass is large,
$\sqrt{\sdark} \sim E_\text{kin}$. %
Overall, we find that the
soft photon approximation describes the $2\to 4$ process well for
small $m_\chi$ in the non-relativistic limit.  However, for
$2m_\chi \sim E_\text{kin}$ or for relativistic
initial states the approximation fails, and it is ultimately related to the
UV-sensitivity of the cross section (see App.~\ref{sec:soft-phot-appr}
for details). Even though calculations simplify considerably in the
soft photon limit, it cannot be applied for the whole $m_{\chi}$-range in
electron bremsstrahlung and we therefore calculate
$\sigma_{2 \rightarrow 4} $ exactly, relegating details of the
calculation to App.~\ref{sec:appdecay-rate-cross}.  However, we will
use the soft photon approximation in its region of validity to
estimate the energy loss from nucleon bremsstrahlung in the next
subsection.

\subsection{Nucleon Bremsstrahlung}
\label{sec:nucleon-bremsstrahlung}

Proton and nuclear bremsstrahlungs are strongly suppressed in
low-temperature environments due to the negligible thermal velocities
of the initial states. %
However, in the interior of a PNS, the typical nucleon
velocity is $v\lesssim 1/3$ and $NN$-bremsstrahlung contributes to the
total energy loss.

The photon that pair-creates the dark states is emitted from the
proton-leg in proton-proton ($pp$) and neutron-proton ($np$)
scattering. Radiating off the neutron-leg through the neutron magnetic
dipole moment is suppressed.  In addition, since $pp$-scattering is
associated with quadrupole radiation, it is suppressed with respect
$np$ scattering by a factor $v^2$~\cite{Rrapaj:2015wgs}. Therefore, we
only consider $np$-scattering in the following.  The interaction of
protons and neutrons is mostly mediated by pions whose mass is of the
order of the average momentum transfer in elastic collisions in PNS,
allowing for a separation of the phase space into an elastic and an
emission piece (see App.~\ref{sec:soft-phot-appr} for details).  The
energy loss rate in the non-relativistic, non-degenerate
limit\footnote{The corrections to Eq.~\eqref{eqn:energy_loss_np} due
  to matter degeneracy in PNS are estimated in
  Ref.~\cite{Rrapaj:2015wgs} to be small ($\approx 30\%$) compared to
  the corrections neglected in the soft-photon approximation, which
  are up to a factor 3 as mentioned in App.~\ref{sec:soft-phot-appr}.}
is then given by~\cite{Rrapaj:2015wgs}
\begin{align} \label{eqn:energy_loss_np}
\begin{split}
	\dot Q_{np}&=
	\frac{n_n n_p}{\sqrt{\pi}\left(m_n T\right)^{3/2}}
	\int_{2m_\chi}^{\infty} \!\!\! dE_\text{kin} \; E_\text{kin}^2 \, e^{-\frac{E_\text{kin} }{T}} 
	\int_{4m_\chi^2}^{E_\text{kin}^2 } \!\! d \sdark \; \\
	&\quad \times
	\sqrt{1- \dfrac{4m_\chi^2}{\sdark}}
	\dfrac{f(\sdark)}{16 \pi^2 \sdark^2} \;
	 \sigma_{np}^T (E_\text{kin}) \; \mathcal{I}_\omega(\sdark) %
	\,,
\end{split}
\end{align}
where $n_n$ and $n_p$ are the neutron and proton number densities,
$m_n$ is the average nucleon mass and $E_\text{kin}$ the available
kinetic center of mass energy. For the elastic $np$-scattering
transport cross section $\sigma_{np}^T$ we use the numerical data from
Fig.~3 in Ref.~\cite{Rrapaj:2015wgs} for $E_\text{kin}\gtrsim 1\,\MeV$
and Fig.~2 in Ref.~\cite{Brown:2018jhj} for smaller energies; the
emission piece $\mathcal{I}_\omega$ is given by
Eq.~\eqref{eqn:emission_piece}.
 
The energy loss rates from nucleon bremsstrahlung are comparable with
the rates from plasmon decay and are depicted for $m_\chi=1 \; \MeV$
as a function of the PNS radius in Fig.~\ref{fig:ELR_SN}.

\section{Constraints on the effective coupling}
\label{sec:constraint}

\begin{figure*}[tb]
\centering
\includegraphics[width=\columnwidth]{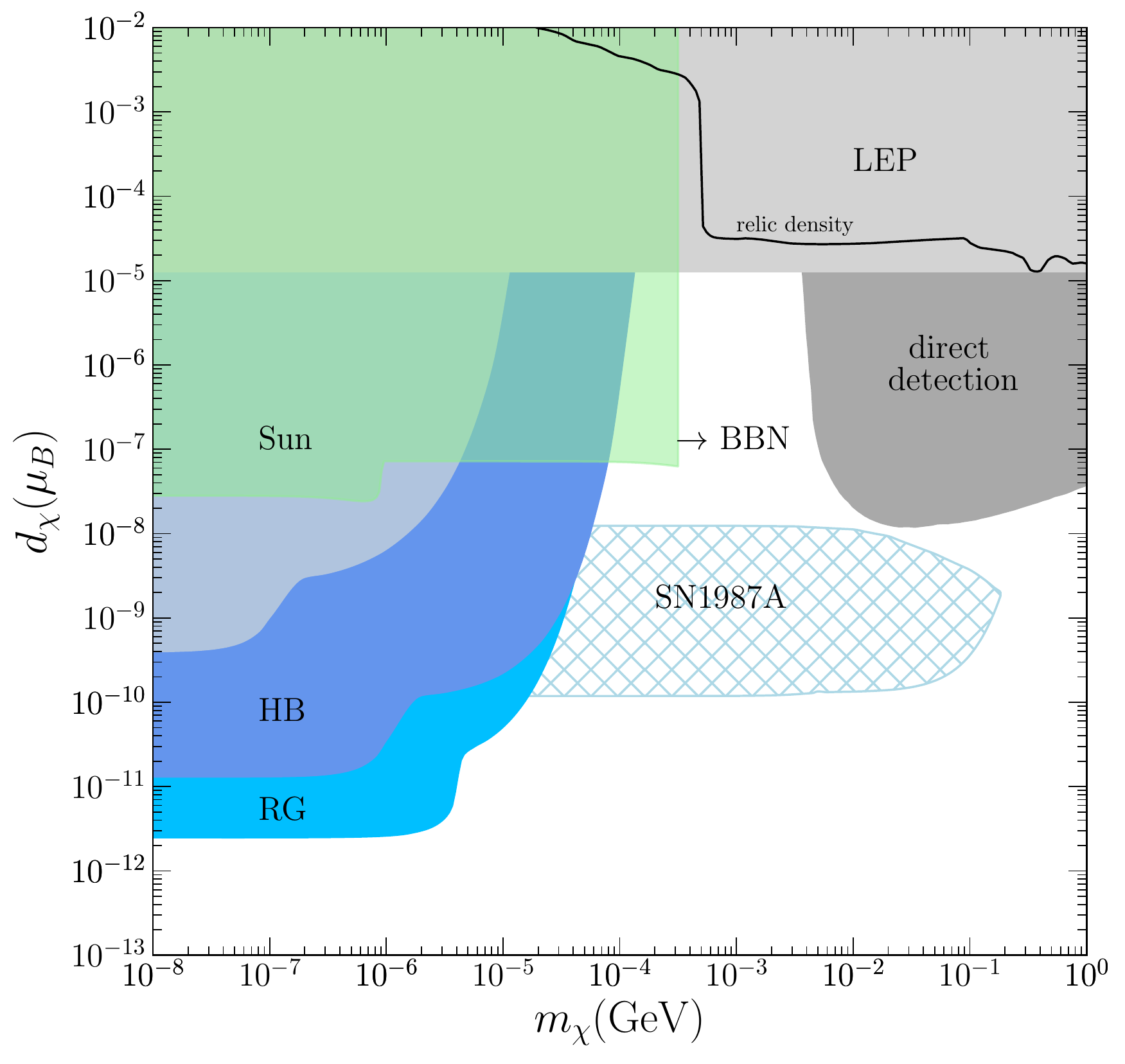}
\includegraphics[width=\columnwidth]{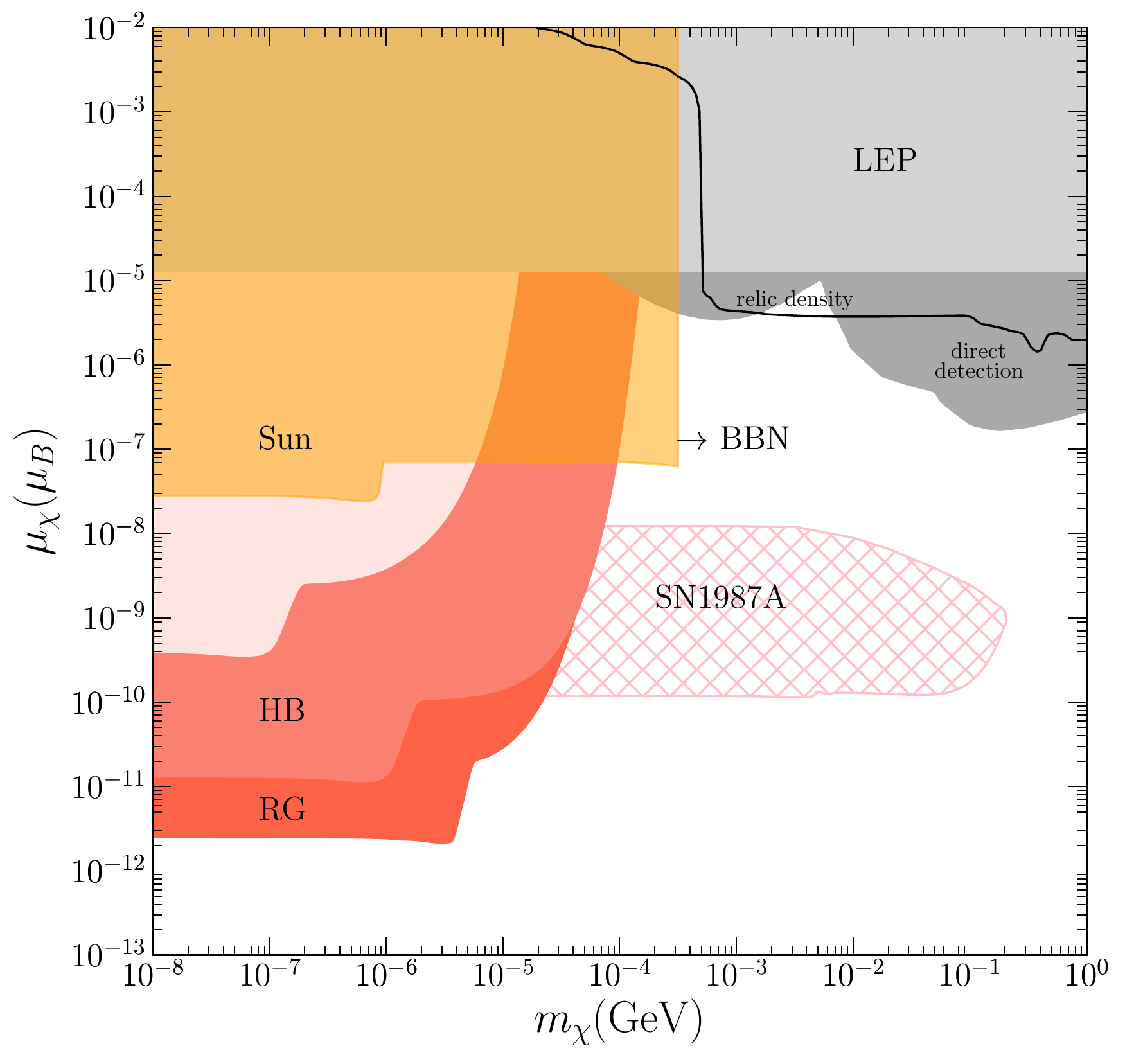}
\caption{Summary of constraints on the EM form factors for dim-5
  operators, \textit{i.e.}~EDM (left) and MDM (right). Colored exclusions are derived in this work. Direct
  detection (only applying to dark matter) and LEP bounds are taken
  from our previous work~\cite{Chu:2018qrm}. On the solid black line
  the thermal freeze-out abundance matches the DM density.
}
  \label{fig:landscapeDim5}
\end{figure*}

\begin{figure*}[tb]
\centering
\includegraphics[width=\columnwidth]{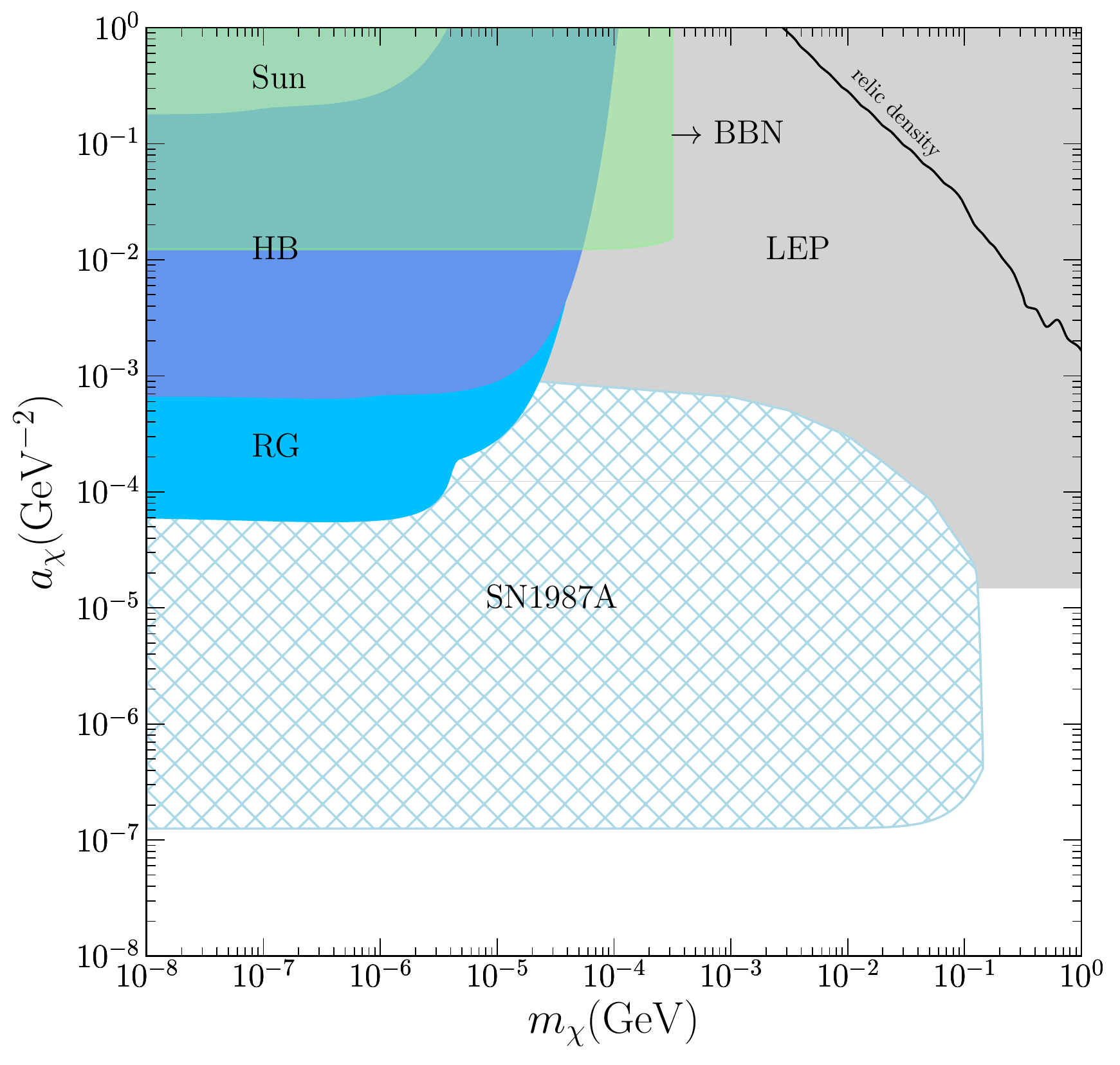}
\includegraphics[width=\columnwidth]{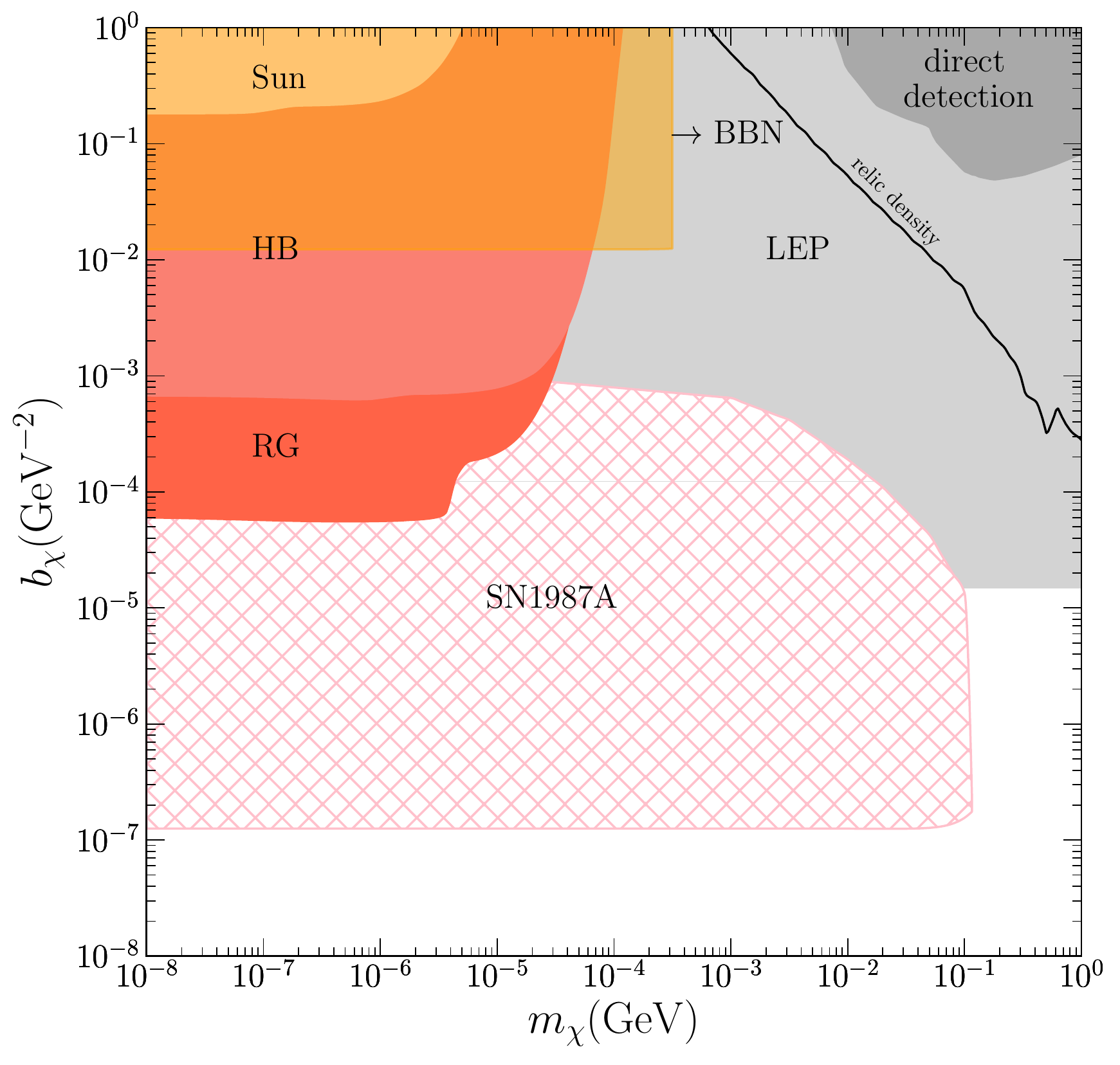}
\caption{Summary of constraints on the EM form factors for dim-6
  operators, \textit{i.e.}~AM (left) and CR (right); labels are the same as in Fig.~\ref{fig:landscapeDim5}.
}
    \label{fig:landscapeDim6}
\end{figure*}

After calculating the energy loss rates induced by those relevant
processes in each stellar environment (e.g. see
Figs.~\ref{fig:ELR_HBsun}-\ref{fig:ELR_SN}), we apply the luminosity
criteria introduced in Sec.~\ref{sec:astr-observ} to obtain the
upper bounds on the EM form-factors of light dark states.

\subsection{Limits from RG, HB, and  Sun}

In this subsection, we derive the constraints coming from HB and RG stars
utilizing the above calculated anomalous energy loss rates.
For HB, we consider a representative star of $0.8\,M_\odot$ and
utilize the stellar profiles for density, temperature and chemical
partition between hydrogen and helium from \cite{Dearborn:1989he,
  Raffelt:1996wa}, reproduced in Fig.~\ref{fig:HBSNprofiles} in
App.~\ref{sec:phot-therm-medi}. The luminosity of its helium-burning core is
$L_{\rm HB}= 20 L_\odot $ to which (\ref{eq:criterionHB}) is then
applied. For RG we use the prescription detailed below
(\ref{eq:criterionRG}): a $0.5\,M_\odot$ helium core with a constant
density of $\rho = 2\times 10^5\, \mathrm{g}/\mathrm{cm}^3$ and a
temperature $T=10^8\,\mathrm{K}$.

For the Sun, we use the standard Solar model
BP05(OP)~\cite{Bahcall:2004pz} to calculate the total power radiated
into $\chi\bar\chi$ which in turn is constrained from
(\ref{eq:criterionSun}). For
bremsstrahlung we take the contribution of electron scattering on H,
$^4$He and other less abundant nuclei ($^3$He, C, N, and O).  For simplicity, we
assume all targets are in a fully ionized state.
We find numerically that the contribution from the second class of elements
contribute $10\%$ of the total energy loss rate from
bremsstrahlung, as the coherent enhancement from atomic charge number $Z$
somewhat compensates for their scarcity in number.

The energy loss rates as a function of fractional stellar radius for
HB (Sun) for all operators considered in (\ref{eq:dim-5}) and
(\ref{eq:dim-6}) are shown in the left (right) panel of
Fig.~\ref{fig:ELR_HBsun} for $m_{\chi} = 10\,\eV$ and
$\mu_{\chi} (\text{or } d_{\chi}) = 10^{-6}\mu_B$ and
$a_{\chi}(\text{or } b_{\chi}) = 0.1 \,\GeV^{-2}$. MDM and EDM as well as AM and
CR lines essentially yield identical results. This is
owed to the fact that production proceeds in the kinematically
unsuppressed region $T\gg m_{\chi}$ for which the energy loss rates
match; the $\gamma^5$ factor discriminating the interactions of same
mass-dimension only plays a role when $\chi$ particles become
non-relativistic, hence close to kinematic endpoints.   As
  can be seen, for dimension-5 operators the decay
  process (dotted lines) dominates over
  bremsstrahlung (dash-dotted lines) and Compton scattering (dashed
  lines) processes in both HB and Sun.  For dimension-6 operators, the
  contribution of Compton scattering is comparable to that of decay
  processes in HB while in the Sun all three processes are of comparable importance.  

  Applying the criteria for the maximum allowable energy loss of
  Sec.~\ref{sec:astr-observ}, we obtain the excluded shaded regions in
  Figs.~\ref{fig:landscapeDim5} and~\ref{fig:landscapeDim6} as
  labeled. The strongest limits are provided by  RG stars. They have
  a higher core temperature, $T=8.6\,\keV$, compared to HB stars or
  the Sun, favoring an emission process that is UV-biased because of
  the considered higher-dimensional operators.
In the low mass region, for $2m_{\chi}< \omega_p$, the limits are
governed by $ \gamma_{\rm T,L}$ decays, and become independent of $\chi$
mass quickly. Once the decay process is kinematically forbidden, the limits become
determined by the bremsstrahlung and Compton scattering processes. As can be seen, the critical values of
 $m_{\chi}$ where this happens for RG, HB, and the Sun are
reflective of the differing core-plasma frequencies
(\ref{eq:omegaP_stars}) of the respective systems. Furthermore, the
mass-dimension~5   constraints on MDM and EDM are practically
identical; differences only appear in the kinematic endpoint region.

\subsection{Limits from SN1987A}

Limits on $\chi$-photon interactions from SN1987A have previously been estimated
in our earlier paper~\cite{Chu:2018qrm}, largely following the approach of 
\cite{Chang:2018rso}, and considering $e^+e^-$ annihilation but with
plasmon decay neglected.  Here we revisit these constraints in light of
a more detailed calculation.
Dark state pairs with mass $m_{\chi}\lesssim 400\,\MeV$ can be
efficiently produced inside PNS, predominantly through $e^+e^-$
annihilation as positrons are thermally supported. Nevertheless, we
will consider all processes in Fig.~\ref{fig:feyn_diag} except for
electron bremsstrahlung as it is significantly weaker than the others;
see Fig.~\ref{fig:ELR_SN} for one example with $m_\chi =1\,$MeV.

When the particles stream freely after production and are hence able
to escape from the PNS core, the limit (\ref{eq:criterionSN})
applies. We set the size of the PNS core to be
$r_\mathrm{core}=15\,\mathrm{km}$ and model the PNS 
from which $\chi\bar \chi$ pairs are emitted using the 
simulation results of a $18 M_\odot$ progenitor
in~\cite{Fischer:2016cyd} (see Fig.~\ref{fig:HBSNprofiles} in App.~\ref{sec:phot-therm-medi}).
Notice that such simulation results are based on an
  artificial neutrino-driven explosion method and should be taken with
  a grain of salt.
We adopt the total energy density $\rho (r)$, temperature $T(r)$ and
electron abundance $Y_e (r)$ profiles at $1 s$ after the core bounce.
The number density of baryons can be computed as
$n_b (r) \simeq \rho (r)/m_p$ and the number density of electrons can
be written as $n_e (r) \simeq n_b (r) Y_e (r)$.
Other quantities such as chemical potential of electrons $\mu_e (r)$,
plasma frequency $\omega_p (r)$ and effective electron mass
$m_e^{\mathrm{eff}} (r)$ are derived from $n_e (r)$ and $T(r)$.
$m_e^{\mathrm{eff}} (r)$ is recursively solved at each radius using
Eq.~\eqref{Eq:effective_me}; see App.~\ref{sec:phot-therm-medi} for
details.
In the calculations that relate to the anomalous emission, $m_e$ is understood to
be $m_e^{\mathrm{eff}} $.

The result is shown by the lower boundary of the region labeled
SN1987A in Figs~\ref{fig:landscapeDim5} and \ref{fig:landscapeDim6}.
Compared to our previous result in~\cite{Chu:2018qrm} where only
$e^+ e^-$ annihilation was taken into account, the constraint for MDM
and EDM is improved. This is traced back to the fact that the energy
loss rate of $\gamma_{\rm T,L}$, nucleon-bremsstrahlung and Compton
scattering for MDM and EDM are comparable to $e^+ e^-$ annihilation
into a $\chi\bar \chi$ pair. For AM and CR, however, the results
from~\cite{Chu:2018qrm} remain largely unchanged, as $\gamma_{\rm T,L}$ decay,
nucleon-bremsstrahlung and Compton scattering are less efficient.

Once the effective coupling becomes large enough, the produced $\chi$
particles will eventually come into thermal equilibrium with SM
particles. Here we follow \cite{Raffelt:2001kv, Keil:2002in} to divide the radial region into 
an inner ``energy sphere'' of radius $r_\text{ES}$, within which light
dark particles are thermalized, and an outer diffusion zone, where the
$\chi$ luminosity gets attenuated by a transmission coefficient
$S_\text{ES}$. Both quantities are obtained from the energy-exchange
and momentum-exchange mean-free-paths, $i.e.$, $\lambda_E(r)$ and
$\lambda_M(r)$, of the $\chi$ particle in the medium~\cite{Raffelt:2001kv}:
\begin{equation}
	\int^{r_\text{inf}}_{r_\text{ES}} {dr \over \sqrt{\lambda_E(r)\lambda_M(r)}} = {2\over 3}\,,
\end{equation}
 and
\begin{equation}
	S_\text{ES} = {1\over 1 + {3\over 4} \int^{r_\text{inf}}_{r_\text{ES}} {dr \over \lambda_M(r)}  }\,,
\end{equation} 
where $r_\text{inf}$ is set to be $35$\,km, beyond which $\chi$
particles free-stream. 

In the PNS environment of interest here, the energy-exchange
mean-free-path is mostly governed by $e$-$\chi$ elastic scattering.
We estimate it at each radius $r$ through
\begin{equation}
\lambda_E (r)= { \langle n_\chi v_\chi  \rangle  \over  \langle  n_e n_\chi \int d\cos\theta  { d\sigma_{e\chi} \over d\cos\theta } v_M \, {|\Delta E |\over T}   \rangle } .%
\end{equation}
Here, $ \langle ... \rangle $ denotes a thermal average over all
participating particles; $n_e$ and $n_\chi$ are the number densities
of electrons and $\chi$ particles which are all assumed to be
in equilibrium.
${ d\sigma_{e\chi} / d\cos\theta }$ is the differential cross section
of $e$-$\chi$ elastic scattering.  The energy-exchange in the
scattering, $|\Delta E|$, is a function of the scattering angle
$\theta$ in the rest frame of the medium. Although not explicitly
written in the expression, we take the Pauli blocking factor for the
final state electron into account.

Meanwhile, the momentum-exchange mean-free-path $\lambda_M(r)$ turns
out to be dominated by $\chi$-nucleon scattering,
\begin{equation}
\lambda_M (r)= { \langle n_\chi v_\chi  \rangle  \over  \sum\limits_{i=n,p}   \langle  n_i n_\chi \int d\cos\theta  { d\sigma_{i\chi} \over d\cos\theta } v_M \, (1-\cos\theta)   \rangle }\,,%
\end{equation}
where $n_{i}$ is the number density of neutrons\,(n) and protons\,(p),
provided by the PNS model~\cite{Fischer:2016cyd}. The differential
scattering cross section with $\chi$ is denoted as
$ { d\sigma_{i\chi} / d\cos\theta } $.  In obtaining our numerical
results, we have neglected the nucleon velocities in the medium frame
for simplicity.  %

As we assume that $\chi$ stay in chemical equilibrium at
$r_\text{ES}$, its blackbody luminosity $L_{\chi}(r_\text{ES})$ is
described by
\begin{equation}
\label{Eq:solve-rd}
L_\chi (r_\text{ES})  =  \dfrac{g_\chi r_\text{ES}^2}{2\pi} \int d|\vec p_\chi| \, \dfrac{|\vec p_\chi|^3}{e^{\sqrt{|\vec p_\chi|^2 + m_\chi^2}/T(r_\text{ES})}+1},
\end{equation}
where $g_\chi = 4 $. %
These $\chi$ particles are emitted towards the exterior from the energy sphere, but they
continue to scatter elastically with the medium inside the diffusion zone. The $\chi$
flux-attenuation during its propagation from $r_\text{ES}$ to
$r_\text{inf}$ can be estimated via
\begin{equation}
	L_\text{inf}= L_\chi (r_\text{ES}) \, S_\text{ES}\,
\end{equation} 
in the diffusion limit~\cite{Raffelt:2001kv}. That is, such
attenuation only relies on the momentum-transfer cross
section. Requiring $L_\text{inf} \le L_\nu$   then leads to   upper
boundaries of SN exclusion limits shown in
Figs.~\ref{fig:landscapeDim5} and \ref{fig:landscapeDim6}.
We note in passing, that the location of the upper boundaries is still
conservative, as additional ``freeze-in'' $\chi$-production for
$r> r_\text{ES}$ has been neglected. With respect to our previous
work~\cite{Chu:2018qrm}, where the transmission ratio $S_{ES}$ was in
practice taken as a simple unit-step function, the upper limits are
improved by up to a factor of two.

\subsection{Related works}
Stellar bounds on the EM properties of light dark states have been
studied in the literature, mostly in the context of EM properties of
eV-scale (SM) neutrinos; see~\cite{Giunti:2014ixa, Tanabashi:2018oca}
and references therein.
In these studies, the mass of neutrino is essentially zero. Therefore,
in the limit $m_\chi \rightarrow 0$ our results can be compared with
previously derived constraints on neutrino EM interactions.

For instance, based on similar energy loss arguments, bounds on the
neutrino MDM have been obtained by calculating the plasmon decay process, from RG as
$\mu_\nu \le ( 2-4) \times 10^{-12}\mu_B$~\cite{Raffelt:1990pj,
  Viaux:2013hca, Arceo-Diaz:2015pva}, from HB as
$\mu_\nu \le (1-3)\times 10^{-11}\mu_B$~\cite{Fukugita:1987uy,
  Raffelt:1987yb}, from the Sun as
$\mu_\nu \le 4\times 10^{-10}\mu_B$~\cite{Raffelt:1999gv}. Indeed, all
these bounds  are in essential agreements with our newly derived ones once the limit $m_\chi \to 0$ is taken.

For higher-dimensional operators, \cite{Grifols:1989vi} estimated the
anomalous energy loss rate in PNS through electron pair annihilation
into light right-handed neutrinos, limiting its charge radius to be
below $3.7\times 10^{-34}$\,cm$^2$, that is
$9.5\times 10^{-7}$\,GeV$^{-2}$, about seven times weaker than the one
presented above. This is partially due to the fact that
\cite{Grifols:1989vi} assumed an one order of magnitude larger
luminosity as the maximum permissible energy loss.

\subsection{Cosmological constraints}
 
Light dark states may lead to extra radiation in the early Universe,
and thus its coupling to the SM bath is constrained by both the
predictions from big-bang nucleosynthesis (BBN) and the observed cosmic
microwave background (CMB). On the one hand, for the mass region
considered here the CMB bounds depend on how it annihilates/decays. 
On the other hand, primordial abundance
measurements of D and $^4$He suggest that extra relativistic 
degrees of freedom need to be less than that of one chiral fermion during the
nucleosynthesis~(see e.g.~\cite{Steigman:2010pa, Mangano:2011ar,
  Hamann:2011ge}). Thus here we require that the Dirac fermion $\chi$
is thermally diminished at $T\sim 100$\,keV, either due to  a feeble EM
form-factor coupling or by a Boltzmann-suppression induced by its mass.

The relevant bounds are also given in Figs.~\ref{fig:landscapeDim5}
and \ref{fig:landscapeDim6}. They only constrain the parameter region
with $m_\chi \ll 1 \,\MeV$. In the same figures, we also show the line
which corresponds to the thermal freeze-out scenario which generates
the observed dark matter abundance, although such scenario has been
excluded by various constraints for this model; see our previous
work~\cite{Chu:2018qrm}. The dominant annihilation channel is into two
photons at $m_\chi <m_e$ and into a pair of electrons at
$m_\chi \ge m_e$, which explains the sharp decrease of the relic
density curve at $m_\chi \sim m_e$ seen in
Fig.~\ref{fig:landscapeDim5}.

\section{Conclusions}
\label{sec:conclusions}

In this paper we explore the sensitivity of stellar systems to neutral dark
states that share higher-dimensional interactions with the SM photon. 
To this end we choose a Dirac fermion
$\chi$ that is coupled to mass dimension 5 MDM and EDM operators with
respective dimensionful coefficients $\mu_{\chi}$ and $d_{\chi}$ and
mass dimension 6 AM and CR operators with respective coefficients
$a_{\chi}$ and $b_{\chi}$. We consider anomalous energy losses from
the interior of RG and HB stars, of the Sun, and of the PNS core of SN1987A. Together with previously derived direct, indirect,
and cosmological limits by us in~\cite{Chu:2018qrm}, this work adds
astrophysical constraints to draw a first comprehensive overview of light
dark states with masses (well) below the GeV-scale and EM moment
interactions.

The thermal environments of stellar interiors significantly affect (or
enable) production processes of $\chi\bar\chi$ pairs. Before breaking
it down to individual contributions, we establish the exact formula,
Eq.~(\ref{eq:generalRate}), for the pair-production rate in leading
order of the dark coupling. The expression factorizes into a piece
that represents the probability to produce an off-shell photon
$\gamma^{*}$ from a SM current, and a piece that describes the
production of the $\chi\bar\chi$ pair from that photon. The former is
proportional to the imaginary parts of the longitudinal and transverse
thermal photon self energies $\imag \Pi_{\rm L,T}$.  The latter are
model-dependent but otherwise universal factors that represent the
choice of interaction, Eq.~(\ref{eqn:f}). The optical theorem then
allows us to identify all major production processes by studying the
contributions to $\imag \Pi_{\rm L,T}$. The approach also allows us to
clarify the role of thermal resonances in these processes,
\textit{i.e.}, the kinematic situation when the pair-producing photon
goes on-shell, $k^2 = \real \Pi_{\rm L,T}$. We find that resonant
production is entirely captured by the decay of transverse and
longitudinal thermal photons or ``plasmons'',
$\gamma_{\rm T,L}\to \chi\bar\chi$.

We compute the rates of $\chi$-pair production and its ensuing energy
loss from plasmon decay and Compton production for all systems. In
addition, we evaluate $eN$ bremsstrahlung for RG, HB and the Sun, and
$e^+e^-$ annihilation and $NN$ bremsstrahlung for SN1978A. For MDM and
EDM interactions, plasmon decay dominates in HB and RG stars and in
the Sun. For the interactions of increased mass dimension, AM and CR,
the Compton (bremsstrahlung) production dominates in HB and RG
(Sun). In PNS core, $e^+e^-$ annihilation dominates the anomalous energy
loss for $r \gtrsim 7\,\rm km$. In the most inner region the
population of positrons becomes extremely Boltzmann suppressed by a
decrease in temperature, and plasmon decay and $np$ bremsstrahlung
take over as the most important production channels. For all processes
we have taken into account all important finite-temperature
effects. Furthermore, in the evaluation of rates, we explicitly avoid
any double counting between plasmon decay and an on-shell emitted
photon in bremsstrahlung and between Compton production and an
on-shell exchanged $t$-channel photon in bremsstrahlung.

The rates when integrated over stellar radius then become subject to
the observationally inferred limits on anomalous energy loss. The
resulting restrictions on the parameter space are found in
Figs.~\ref{fig:landscapeDim5} and~\ref{fig:landscapeDim6}. In the
kinematically unrestricted regime $m_{\chi} \lesssim 1\,\keV$, the
stellar limits are dominated by RG with
$\mu_{\chi}, d_{\chi} \leq 2\times 10^{-12}\mu_B$ and
$a_{\chi}, b_{\chi} \leq 6\times 10^{-5}\, \GeV^{-2}$. All
interactions are additionally constrained from SN1987A, in the windows
$ 10^{-10}\mu_B \leq \mu_{\chi}, d_{\chi} \leq 10^{-8}\,\mu_B$ and
$ 10^{-7}\,\GeV^{-2} \leq a_{\chi}, b_{\chi} \leq 10^{-3}\,\GeV^{-2}$
for $m_{\chi} \lesssim 10\,\MeV$. The SN constraining region is
bounded from above by the trapping of $\chi$ particles, which we
evaluate in some detail.
The presented astrophysical constraints add to a program that we have
started in \cite{Chu:2018qrm} and that aims at charting out the
experimental and observational sensitivity to effective dark
state-photon interactions. The stellar constraints on anomalous energy
loss derived in this work yield the most important limits on the
existence of effective dark sector-photon interactions for
$\chi$-particles below the MeV-scale.

\vspace{.3cm}

\paragraph*{Acknowledgments}
 The authors are supported by the New Frontiers program
  of the Austrian Academy of Sciences. JLK and LS are supported by the Austrian
  Science Fund FWF under the Doctoral Program W1252-N27 Particles and
  Interactions. We acknowledge the use of computer packages for
  algebraic calculations~\cite{Mertig:1990an,Shtabovenko:2016sxi}.

\appendix

\section{Photons in a thermal medium}
\label{sec:phot-therm-medi}

The processes depicted in Fig.~\ref{fig:feyn_diag} are
fundamentally affected (or enabled) by the in-medium modified photon
dispersion.  Here we collect
the central results that go into the computation of the energy loss
rates (our convention largely follows \cite{Raffelt:1996wa}). The central quantity measuring the strength of the
medium-effect is the plasma frequency $\omega_p$, obtained through
\begin{equation}
\label{Eq:omegap_general}
\omega_p^2 = \dfrac{4\alpha}{\pi} \int_0^\infty dp \dfrac{p^2}{E} \left(1- \dfrac{1}{3} v^2 \right) \left(f_{e^-} + f_{e^+} \right),
\end{equation}
where $v = p/E$ is the velocity of electrons or positrons, and
$f_{e^-}$ and $f_{e^+}$ are their respective Fermi-Dirac
distributions, $f_{e^\pm} = [e^{(E\pm\mu_e)/T}+1]^{-1}.$

Equation (\ref{Eq:omegap_general}) takes on the following analytic forms in the classical, degenerate and relativistic limit respectively,
\begin{equation}
\omega_p^2 \simeq
 \begin{cases}
    \dfrac{4\pi \alpha n_e}{m_e} \left( 1 - \dfrac{5}{2} \dfrac{T}{m_e}\right)& \text{classical}  \\
    \dfrac{4\pi \alpha n_e}{E_F} = \dfrac{4\alpha}{3\pi} p_F^2 v_F & \text{degenerate}  \\
\dfrac{4\alpha}{3\pi} \left( \mu_e^2 + \dfrac{1}{3} \pi^2 T^2\right)& \text{relativistic}   \\ 
  \end{cases}\,,
\end{equation}
where $\alpha$ is the fine-structure constant, $n_e$ is the number
density of electrons, $p_F = (3\pi^2 n_e)^{1/3}$ is the Fermi
momentum, $E_F = \sqrt{m_e^2 + p_F^2}$ is the Fermi energy and
$v_F = p_F/E_F$ is the Fermi velocity.
Here ``classical'' refers to a non-relativistic ($T \ll m_e$) and
non-degenerate ($T \gg \mu_e - m_e$) plasma. 

The PNS core of a SN is both in a relativistic and
degenerate regime and we find that the relativistic limit above
yields a better fit to the general form of $\omega_p$ in
(\ref{Eq:omegap_general}) than the degenerate limit; the latter
exhibits a $10\%$ deviation. The core of a RG star is
non-relativistic but degenerate whereas HB stars and the Sun
are well described by the classical limit.  In our numerical
calculations, we adopt $\omega_p$ computed from
Eq.~\eqref{Eq:omegap_general}, avoiding any ambiguities of taking
limiting cases.  Representative values of $\omega_p$ at the cores of
all stellar objects are summarized as
\begin{equation}
    \label{eq:omegaP_stars}
  \omega_p \sim
  \begin{cases}
    0.3 \, \keV & \text{Sun's core} \\
    2.6 \,\keV & \text{HB's core} \\
    8.6 \, \keV & \text{RG's core}   \\ 
    17.6 \, \MeV & \text{SN's core} \\
  \end{cases}\,. 
\end{equation}
The computation of most of the processes requires the in-medium photon
propagator.  Picking Coulomb gauge, for a photon carrying 4-momentum $k = (\omega, \vec k)$, the latter divides into
longitudinal (L) and transverse (T) parts~\cite{Braaten:1993jw}, 
\begin{equation}
\label{Eq:Propagator_Coulomb}
\begin{split}
	D_{00} &= \dfrac{k^2}{|\vec k|^2(k^2 -\Pi_\mathrm{L})}\,g_{00} \,,\\
	D_{ij} &= \dfrac{1}{k^2- \Pi_\mathrm{T} }\left(\delta_{ij}-\dfrac{k_i k_j}{|\vec k|^2}\right)\,,
\end{split}
\end{equation}
where $k_i$ is the Cartesian component of the photon three-momentum
(magnitude $|\vec k|$).
Using~\cite{Braaten:1993jw} and adopting the conventions
of~\cite{Raffelt:1996wa, An:2013yfc}, the real part of the polarization functions
$\Pi_{\rm T, L}$ in the rest frame of the (isotropic) thermal bath
reads,
\begin{align}
\label{Eq:Pol_tensor}
\begin{split}
  \mathrm{Re}\, \Pi_\mathrm{T}&= \dfrac{3\omega^2}{2v_*^2 |\vec k|^2} \omega_p^2
  \left[ 1- \dfrac{\omega^2-v_*^2 |\vec k|^2}{2\omega v_*  |\vec k|}
    \ln \dfrac{\omega + v_* |\vec k|}{\omega-v_* |\vec k|} \right], \\
  \mathrm{Re}\, \Pi_\mathrm{L} &=  3 \omega_p^2
  \left(\dfrac{\omega^2 - |\vec k|^2}{v_*^2 |\vec k|^2} \right) \left[ \dfrac{\omega}{2v_* |\vec k|} \ln \dfrac{\omega + v_* |\vec k|}{\omega - v_*|\vec k|} -1 \right].
\end{split}
\end{align}
The full expressions for the dispersion relations $k^2 -\Pi_\mathrm{L, T}=0$ then relate the
energies  of an on-shell photon, $\omega_{\rm T, L}$, to its 
momentum $\vec k$ to order $\alpha$ \cite{Braaten:1993jw}, via  
\begin{equation}
\label{Eq:dispersion_relation}
\begin{split}
\omega_\mathrm{T}^2 &= |\vec k|^2 +\omega_p^2 \dfrac{3 \omega_\mathrm{T}^2}{2v_*^2 |\vec k|^2} \left[ 1- \dfrac{\omega_\mathrm{T}^2 - v_*^2 |\vec k|^2}{2\omega_\mathrm{T} v_* |\vec k|} \ln \dfrac{\omega_\mathrm{T} + v_* |\vec k|}{\omega_\mathrm{T}- v_* |\vec k|} \right], \\ 
\omega_\mathrm{L}^2 &= \omega_p^2 \dfrac{3 \omega_\mathrm{L}^2}{v_*^2 |\vec k|^2} \left[ \dfrac{\omega_\mathrm{L}}{2v_* |\vec k|} \ln \dfrac{\omega_\mathrm{L} + v_* |\vec k|}{\omega_\mathrm{L} - v_* |\vec k|} -1 \right].
\end{split}
\end{equation}
Equations~\eqref{Eq:dispersion_relation} are also valid to order
$|\vec k|^2$ at small $|\vec k|$ for all temperatures and electron
number densities. Throughout the paper, we always use
$\omega_{\rm T, L}$, as functions of $|\vec k|$, to denote the energy
of an on-shell thermal photon, which satisfies
Equations~\eqref{Eq:dispersion_relation}, and use $\omega$ for
off-shell photons.

Longitudinal photons are populated up to a wavenumber $k_{max}$, beyond which the
longitudinal dispersion relation crosses the light cone and L-modes
become damped, with
\begin{equation}
\label{Eq:kmax}
k_{max} = \left[\dfrac{3}{v_*^2} \left( \dfrac{1}{2 v_*} \ln \dfrac{1+ v_*}{1- v_*} - 1 \right) \right]^{1/2} \omega_p \,,
\end{equation}
and in the relativistic limit $k_{max} \to \infty$. In these equations,
the mobility of charges is captured by the typical velocity of electrons,
$v_* \equiv \omega_1 /\omega_p$, where
\begin{equation}
\omega_1^2 = \dfrac{4\alpha}{\pi} \int_0^\infty dp \dfrac{p^2}{E} \left(\dfrac{5}{3} v^2 - v^4 \right) \left(f_{e^-} + f_{e^+} \right).
\end{equation}
In the three limits mentioned previously, $v_*$ can be approximated as
\begin{equation}
 v_* \simeq
  \begin{cases}
  \sqrt{5T / m_e }& \text{classical} \\
    v_F & \text{degenerate} \\
 1& \text{relativistic}   \\ 
\end{cases}.
\end{equation}
Finally, as alluded to in the main text, the processes we consider are
non-resonant in the photon exchange and
$\mathrm{Im}\, \Pi_{\mathrm{T},\mathrm{L}}$ can be neglected
throughout. %

In turn, the computation of in-medium photon decay,
i.e.~process~(\ref{eq:plasmon}a), requires the description of external
in-medium photon states. For propagation in the $z$-direction,
i.e.  $k_x=k_y=0$, the transverse and longitudinal
polarization vectors are given by
\begin{subequations}
  \label{eq:polvectors}
\begin{equation}
  \epsilon^\mu_\mathrm{T} = (0 ,1 (0) ,0 (1), 0)
  ,\quad \epsilon^\mu_\mathrm{L} = \dfrac{1}{\sqrt{\omega_L^2 -|\vec k|^2}} (|\vec k|, 0 ,0 ,\omega_L).
\end{equation}
\end{subequations}
In all cases $\epsilon^\mu \epsilon_\mu  =-1$ and $\epsilon^\mu k_\mu=0$.

Furthermore, the in-medium  coupling of the photon to the
EM current is modified by the vertex renormalization
constants
$Z_{\mathrm{T},\mathrm{L}}\equiv (1-\partial
\Pi_{\mathrm{T},\mathrm{L}}/\partial \omega_{\rm T, L }^2)^{-1}$.  For the
convention adopted here, they are equivalent to the ones given
in~\cite{Raffelt:1996wa}%
\begin{equation}
\label{Eq:vertex_renorm}
\begin{gathered}
Z_\mathrm{T} = \dfrac{2\omega_\mathrm{T}^2 (\omega_\mathrm{T}^2 - v_*^2 |\vec k|^2)}{3\omega_p^2 \omega_\mathrm{T}^2 +(\omega_\mathrm{T}^2 +|\vec k|^2)(\omega_\mathrm{T}^2 - v_*^2 |\vec k|^2) -2\omega_\mathrm{T}^2 (\omega_\mathrm{T}^2 -|\vec k|^2)},  \\
Z_\mathrm{L} = \dfrac{2(\omega_\mathrm{L}^2 -v_*^2 |\vec k|^2)}{3\omega_p^2 - (\omega_\mathrm{L}^2 - v_*^2 |\vec k|^2)} \frac{\omega_L^2}{\omega_L^2-|\vec k|^2}\,,
\end{gathered}
\end{equation}
These factors are attached to each zero-temperature vertex factor involving an
external photon state. For internal photons, this effect is
already accounted for in the momentum-dependent self-energy
$\Pi_{\rm T, L} (\omega, \vec k)$. %

\begin{figure*}[tb]
\centering
\includegraphics[width=0.95\columnwidth]{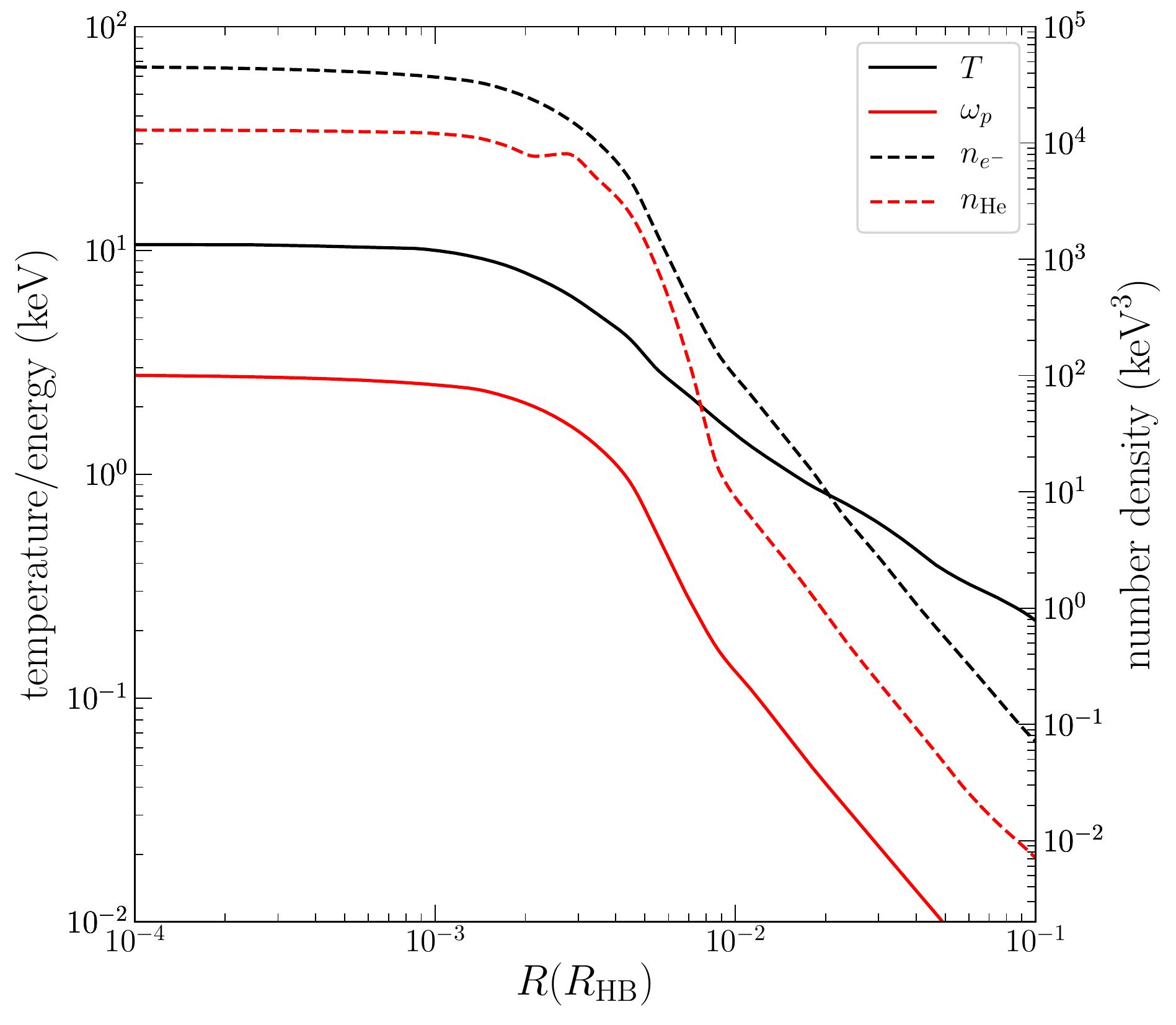}\hfill
\includegraphics[width=0.92\columnwidth]{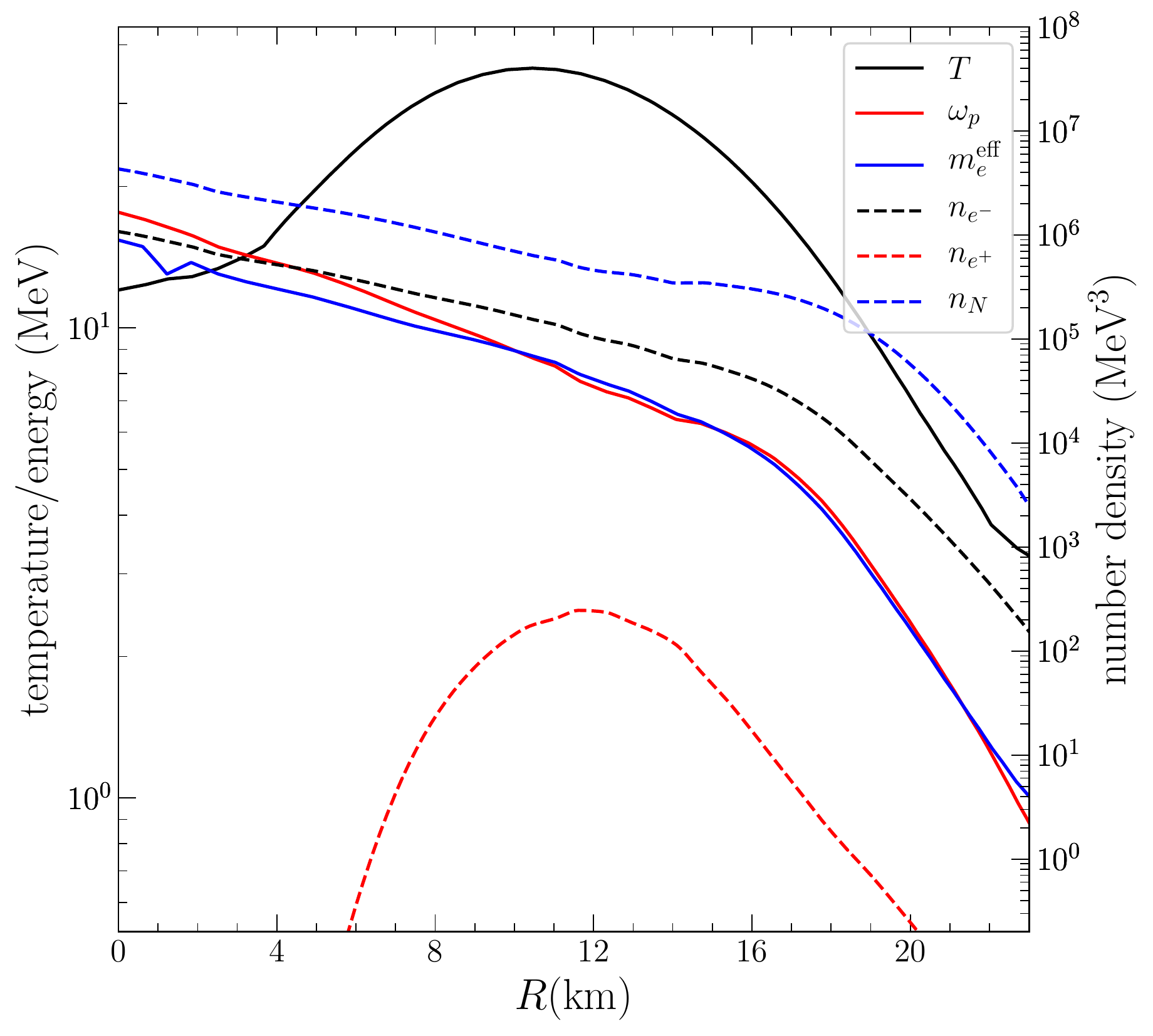}
\caption{Reproduced profiles of a representative $0.8M_{\odot}$ HB
  star~\cite{Dearborn:1989he} (left) and of a PNS of a $18 M_{\odot}$
  progenitor~\cite{Fischer:2016cyd} (right) that are adopted in our
  work. In each panel, the left vertical axis corresponds to the
  values of temperature and plasma frequency (solid lines) at each
  radius, in units of keV(HB) or MeV(PNS), and the right vertical axis
  gives the number densities (dashed lines) of each particle species,
  in ${\rm keV}^3$(HB) or ${\rm MeV}^3$(PNS). For SN, the effective
  electron mass $m_e^{\rm eff}$ is also displayed.}
  \label{fig:HBSNprofiles}
\end{figure*}

For thermal corrections to the electron mass which is relevant for PNS,
we closely follow \cite{Hardy:1998if}. For an electron with a general
4-momentum $p = (E, \vec p)$ in a neutral medium where the positron
number density is negligible, we first introduce the four functions
below
\begin{eqnarray}
A_e &=&{-\alpha  \over 4\pi |\vec p|}\int_0^\infty 
\!\!\!\!\!\! d q 
{q \,f_{e^-} (q)  \over \sqrt{q^2 + m_e^2} }\left[ 4 |\vec p| q   - (p^2+m_e^2) L_2 \right] ,\\
C_e &=&{ \alpha m_e \over \pi |\vec p|} \int_0^\infty 
\!\!\!\!\!\! d q 
{q \,f_{e^-} (q)  \over \sqrt{q^2 + m_e^2} } \, \left( - L_2\right),\\
A_\gamma &=&{-\alpha  \over 4\pi |\vec p|}\int_0^\infty 
\!\!\!\!\!\! d q 
f_\gamma (q) \left[ 8 |\vec p| q + (p^2+m_e^2)(L_3-L_4) \right] ,~~~~~~\\
C_\gamma &=&{ \alpha m_e \over \pi |\vec p|} \int_0^\infty 
\!\!\!\!\!\! d q  f_\gamma (q) \left(L_3 -L_4 \right)	 ,
\end{eqnarray}
where $q$ here is the absolute value of the 3-momentum of medium particles (electron, photon) that is integrated over and $L_{1, 2,3,4}$ are functions of $q$ in terms of 
\begin{align}
L_1(q) &=\ln\left[ {2 (E \sqrt{q^2 + m_e^2}  + |\vec p | q ) - p^2 - m_e^2 \over 2 (E \sqrt{q^2 + m_e^2} - |\vec p | q) - p^2  - m_e^2}\right], \\
L_2(q) &=\ln\left[ {2 (E \sqrt{q^2 + m_e^2}  + |\vec p | q )+ p^2 + m_e^2 \over 2 (E \sqrt{q^2 + m_e^2} - |\vec p | q) + p^2 + m_e^2}\right], \\
L_3(q) &=\ln\left[ {2 (E q + |\vec p | q )+ p^2 - m_e^2 \over 2 (E q- |\vec p | q) + p^2 - m_e^2}\right],\\
L_4(q) &=\ln\left[ {2 (E q + |\vec p | q ) - p^2 + m_e^2 \over 2 (E q- |\vec p | q) - p^2 + m_e^2}\right].
\end{align}
Here $m_e$ is the zero-temperature mass of electron, 0.511\,MeV, while
$f_\gamma (p)$ and $f_{e^-} (p)$ give the thermal momentum
distribution functions of photon and electron (per degree of
freedom). We have set $f_{e^+} (p) =0$ in the above
equations. %
In the end, we take the approximation made in \cite{Donoghue:1983qx} to obtain that 
\begin{align}
	\label{Eq:effective_me}
m_e^\text{eff} (p) = \sqrt{ m_e^2  -2 (A_\gamma + A_{e}) - 2m_e (C_\gamma + C_{e}) }\, . 
\end{align}
We have neglected thermal corrections to $\chi$
states. In the phenomenologically relevant regime, their coupling to the thermal bath is very weak.

Finally, we have reproduced the profiles of the HB model from \cite{Dearborn:1989he}, and PNS model from \cite{Fischer:2016cyd}, adopted in this work, as shown in Fig.~\ref{fig:HBSNprofiles}, where   neutrality and $\mu_{e^-}+ \mu_{e^+}=0$ at each radius have been taken for granted for the PNS profile.

\section{Decay rate and cross section calculations}
\label{sec:appdecay-rate-cross}

In this appendix we collect some further details that enter the
calculation of the $\chi\bar\chi$ production cross sections found in
Sec.~\ref{sec:produc-DM}. 

\subsection{Full expression  of \texorpdfstring{\boldmath$\chi$}{chi}-pair production rate}

For any process that produces a $\chi\bar \chi$ pair through a photon
propagator of 4-momentum $k = (\omega, \vec k)$, its  spin-summed squared matrix
element can be written in terms of %
\begin{equation}
\sum_{\rm spins}	|\mathcal{M} |^2 =  D_{\mu\nu}(k) D^{*}_{\rho\sigma}(k)   \mathcal{T}_\text{SM}^{\mu\rho}  \mathcal{T}_\chi^{\nu\sigma},
\end{equation}
where the in-medium photon propagator $D^{\mu\nu}$ is given by
\eqref{Eq:Propagator_Coulomb}, while $\mathcal{T}^{\mu\rho}_\text{SM}$
and $\mathcal{T}^{\nu\sigma}_\chi$ represent the corresponding squared
matrix elements of the SM current, \emph{i.e.},
$\text{SM} \to \gamma^*(k) ~( + \text{SM}')$, and the dark current,
\emph{i.e.}, $\gamma^*(k) \to \chi (p_\chi) + \bar\chi(p_{\bar\chi})$, of which
the latter is given by
\begin{equation}
\label{Eq:darkTrace}
\mathcal{T}^{\nu\sigma}_\chi = \mathrm{Tr}[(\slashed p_\chi +m_\chi) \Gamma^\nu (k) (\slashed p_{\bar\chi}-m_\chi) \Gamma^\sigma (- k)].
\end{equation}
The vertex factors $\Gamma^{\nu}(k)$ are derived from the Lagrangians
(\ref{eq:dim-5}) and (\ref{eq:dim-6}) through the usual prescription
of obtaining Feynman rules; see \cite{Chu:2018qrm} for the explicit
expressions.  Generalizing Eq.~(5.156) of \cite{Bellac:2011kqa} yields
an expression for the exact $\chi\bar \chi$ differential production rate per volume,
\begin{equation}\label{eq:generalproduction}
  { d\dot{N}_{\chi} \over d^4k} = { 1 \over (2\pi)^4 }  D_{\mu\nu}(k) D^{*}_{\rho\sigma}(k)
  \left({ 2 \imag{\Pi}^{\mu\rho}(k)\over e^{\omega/T}-1 }\right)   I_\chi^{\nu\sigma}\,,
\end{equation}
where $\imag\Pi^{\mu\rho}$ is the imaginary part of the thermal photon
self-energy induced by all possible SM currents. In the medium it is
decomposed into longitudinal and transverse components,
$\imag\Pi_{L,T}$, as shown in Eq.~(\ref{eq:PIdecomp}) in the main text.
The factor $I^{\nu\sigma}$ is the  2-body final state integrated over its phase space,
\begin{equation}
I_\chi^{\nu\sigma} =  \int d\Pi_{i=\chi,\bar\chi}  (2\pi)^4 \delta^4 (k-p_\chi -p_{\bar\chi}) \mathcal{T}^{\nu\sigma}_\chi\,,
\end{equation}
where $d\Pi_i = \prod_i d^3\vec p_i(2\pi)^{-3}(2E_i)^{-1}$, as
mentioned in the main text. The integration can be executed in an
arbitrary frame, and in particular in the rest frame of the thermal
bath by adopting Lenard's formula~\cite{PhysRev.90.968}, generalized to
massive final states. We find,
\begin{equation}
\label{Eq:Lenard_massive}
\begin{split}
 &\int d\Pi_{i=\chi,\bar\chi} (2\pi)^4  \delta^4 (k-p_\chi -p_{\bar\chi}) p_\chi^\mu p_{\bar\chi}^\nu \\
 &=\dfrac{1}{96\pi}(Ak^2 g^{\mu\nu} +2B k^\mu k^\nu)\,,
\end{split}
\end{equation}
where the coefficients $A$ and $B$ are given by
\begin{equation*}
\begin{split}
A & = %
\left( 1-\dfrac{4m_\chi^2}{\sdark}\right)^{3/2},\quad 
B  =\sqrt{1-\dfrac{4m_\chi^2}{\sdark}} \left( 1+\dfrac{2m_\chi^2}{\sdark}\right),
\end{split}
\end{equation*}
with $\sdark=k^2$. In terms of the functions $f(\sdark)$ defined in
(\ref{eqn:f}), the factor $I^{\nu\sigma}$ is then explicitly given by,
\begin{equation}
\label{Eq:PhaseI}
I^{\nu\sigma} = \dfrac{1}{8\pi} \sqrt{1-\dfrac{4 m_\chi^2}{\sdark}} f(\sdark) \left(-g^{\nu\sigma} + \dfrac{k^{\nu}k^{\sigma}}{\sdark} \right)\,.
\end{equation}

Putting all of the above together, we obtain the differential production rate per volume
(\ref{eq:generalRate}) found in the main text which we repeat here for convenience,
\begin{equation}\label{eq:appgeneralRate}
\begin{split}
	 { d\dot{N}_{\chi} \over d^4k}  =&
  \dfrac{1}{64 \pi^5}
  \left[-{\imag\Pi_\mathrm{L}(k) \over |\sdark-\Pi_\mathrm{L}|^2} - {2 \imag\Pi_\mathrm{T} (k)\over |\sdark-\Pi_\mathrm{T}|^2 } \right]  \\
  &\times
  f_B(\omega) \; f(\sdark) \sqrt{1-\frac{4 m_\chi^2}{\sdark}} \, ,
\end{split}
\end{equation}
where $f_B(\omega) = (e^{\omega/T}-1 )^{-1}$ is the Bose-Einstein
momentum distribution of thermal bosons.  During the derivation we
have used  that
$$ -g^{\nu\sigma} + \dfrac{k^{\nu}k^{\sigma}}{\sdark} =
\epsilon_{\mathrm{T},1}^{\nu} \epsilon_{\mathrm{T},1}^{\sigma} + \epsilon_{\mathrm{T},2}^{\nu}
\epsilon_{\mathrm{T},2}^{\sigma} + \epsilon_\mathrm{L}^{\nu} \epsilon_\mathrm{L}^{\sigma}\,.$$

\subsection{Leading contributions to \texorpdfstring{\boldmath$\imag\Pi_{T,L}$}{ImPi}}
\label{sec:lead-contr-imagp}

We now demonstrate that (\ref{eq:generalRate}) or, equivalently,
(\ref{eq:generalproduction}) contain the leading production mechanisms
considered in this paper. In particular we clarify the role of
resonances, and that they are accounted for by the process
$\gamma_{\rm T,L}\to \chi\bar\chi$. To this end, we isolate the pole
contribution to the total production rate, \textit{i.e.}~the case
$\sdark = \real\Pi_{\rm L,T}$. To this end, we adopt the narrow width
approximation,
\begin{equation}\label{eq:NWA}
	\lim_{\imag\Pi_{\rm L,T} \to 0}{-  \imag\Pi_{\rm L,T} (k) \over \pi |\sdark -\Pi_{\rm L,T}|^2}  =\delta(\sdark-\real\Pi_{\rm L,T}) \,,
\end{equation}
where  $\imag\Pi_{\rm L,T} < 0$.
Then noting that $\real\Pi_{\rm L,T}$ is also a function of $\sdark$
and writing $d^4k = d^3\vec k d\sdark /(2\omega) $ yield
\begin{eqnarray}\label{eq:decaygeneral}
  {\dot{N}_{\chi}^{\rm T, L} } & =& g_{\rm T, L}\! \int \!\! {d^3\vec k \over  (2\pi)^3 } f_B(\omega_{\rm T, L}) \notag \\ &&\times \left[ \dfrac{ Z_{\rm T, L} f(\real\Pi_{\rm T, L})  }{16 \pi\omega_{\rm T, L} }
  \sqrt{1-\frac{4 m_\chi^2}{\real\Pi_{\rm T, L}}}\right],
\end{eqnarray}
where $g_{\rm T}=2$ and $g_{\rm L}=1$, counting the degrees of freedom
of the photon modes. Now both $\omega_{\rm T, L}$ and
$\real\Pi_{\rm L,T}$ need to satisfy the photon dispersion relation
with a 3-momentum $\vec k$ due to the $\delta$-function above. As will
be calculated below and given explicitly in \eqref{eq:decayT}, the
term in the parentheses is exactly the decay rate of
$\gamma_{\rm L,T}$ into $\chi$ pairs.%

In a next step we further verify that the contribution of the one
electron loop (OEL) to $\imag\Pi_{\mu\rho}$ induces the production
rate of $\chi$ from electron pair annihilation\footnote{The contribution of the two and three electron loops to $\imag\Pi_{\mu\rho}$ correspond to Compton scattering and bremsstrahlung, respectively.},   process
(\ref{eq:annihilation}b).
In this case, it is easier to start with \eqref{eq:generalproduction},
where according to the in-medium optical theorem (see
Fig.~\ref{fig:optical}) we may write 
\begin{eqnarray}
  \left. 2 \imag\Pi^{\mu\rho}\right|_\text{OEL} &=& \int d\Pi_{i=1,2} \mathcal{T}_e^{\mu\rho} (1- f_{e^-} - f_{e^+})
 \notag\\ && \times  (2\pi)^4\delta^4(k- p_1 -p_2)  \,,
	 \end{eqnarray} 
         where
         $\mathcal{T}_e^{\mu\rho} = \mathcal{M}_{\gamma^*\to
           e^+e^-}^\mu \mathcal{M}_{e^+e^-\to \gamma^*}^{\rho}$, and
         $f_{e^\mp}$ gives the momentum distribution function of
         $e^- (p_1)$, $e^+ (p_2)$ per degree of freedom as defined above. Moreover,  terms that are kinetically forbidden for $k^2>0$ have been neglected~\cite{Weldon:1983jn}. The presence of
         $(1-f_{e^-} - f_{e^+})$ is due to quantum statistics, and would
         disappear for classical particles. Substituting this
         expression into \eqref{eq:generalproduction} gives
\begin{eqnarray}
 \left. { d\dot{N}_{\chi} \over d^4k}\right|_\text{OEL}& =&  \int d\Pi_{i=1,2,\chi,\bar\chi} |\mathcal{M}_{\mathrm{ann}}|^2
      (1- f_{e^-} - f_{e^+})  f_B(\omega)\notag \\ & \times & (2\pi)^4\delta^4(k - p_1 -p_2)\delta^4(k - p_\chi -p_{\bar\chi}) \,.
\end{eqnarray}
Then for the Fermi-Dirac distribution function $f_{e^\pm}$ and the
Bose-Einstein distribution $f_B(\omega)$ with the energy conservation
$E_1 + E_2 = \omega $, there exists the relation 
\begin{equation}
	\left({f_{e^-} \over 1 - f_{e^-} }\right) \, \left({f_{e^+} \over 1 - f_{e^+} }\right)   = {f_B (\omega) \over 1+ f_B(\omega)}\, 
\end{equation}
allowing us to rewrite the number production rate per volume above as 
\begin{equation}
\begin{split}
	\left. \dot{N}_{\chi} \right|_{\text{OEL}} =& \int d\Pi_{i=1,2,\chi,\bar\chi} |\mathcal{M}_{\mathrm{ann}}|^2 f_{e^-}  f_{e^+}   \\
	 &\times   (2\pi)^4\delta^4( p_1  + p_2 - p_\chi -p_{\bar\chi}) \,.
\end{split}
\end{equation}
after integrating over $d^4k$ on both sides. The last expression
transforms precisely to the corresponding energy loss rate
\eqref{eq:Qann}, once both the energy-loss factor $(E_1 + E_2)$ and
fermionic degrees of freedom $f_{e^\pm}$ are taken in account.

\subsection{\texorpdfstring{\boldmath$\gamma_{T,L}$}{gamma} decay to dark states}

In the following appendices we calculate the leading processes in the
usual Feynman-diagrammatic approach using tree-level perturbation
theory augmented by the thermal corrections outlined in Appendix.~\ref{sec:phot-therm-medi}. 
 The decay of a transverse or longitudinal
photon of 4-momentum $k$ to a pair of dark states
$\bar\chi (p_{\bar\chi}) + \chi(p_\chi) $ is described by the
spin-summed squared matrix element
\begin{equation}
\sum\limits_{\rm spins} |\mathcal{M}_{\mathrm{T},\mathrm{L}}|^2 = Z_{\mathrm{T},\mathrm{L}}\epsilon_\mu(k) \epsilon_\nu^* (k) \mathcal{T}_\chi^{\mu\nu}
\end{equation}
where $Z_{\mathrm{T},\mathrm{L}}$ is the vertex renormalization factor
in~\eqref{Eq:vertex_renorm}, $ \epsilon_{\mu}$ is the photon
polarization vector and $\mathcal{T}_\chi^{\mu\nu}$ is given
in~\eqref{Eq:darkTrace}.
The decay rate is given by the phase-space integral, 
\begin{equation}
  \Gamma_{\mathrm{T},\mathrm{L}} = \!\int \!\! d\Pi_{i=\chi,{\bar\chi}}  (2\pi)^4
  \delta^4 (k-p_\chi -p_{\bar\chi}) \dfrac{1}{2\omega_{\rm T, L}} \sum\limits_{\rm spins} |\mathcal{M}_{\mathrm{T},\mathrm{L}}|^2,
\end{equation}
where $\omega_{\rm T, L}$ is the energy of the external transverse or longitudinal photon. %
It is useful to employ~\eqref{Eq:Lenard_massive}.
In terms of $I^{\nu\sigma}$ defined in~\eqref{Eq:PhaseI}, we can write
the decay rate as
\begin{equation}
\Gamma_{\mathrm{T},\mathrm{L}} = \dfrac{1}{2\omega_{\rm T, L}} Z_{\mathrm{T},\mathrm{L}} \epsilon_\mu (k) \epsilon^*_\nu (k) I^{\mu\nu}.
\end{equation}
The explicit expression, given by \eqref{eq:decayT} in the main text,
is then found by using the expressions~(\ref{eq:polvectors}) for the
polarization vectors when the initial state propagates in positive
$z$-direction, \textit{i.e.}~for $k^\mu = (\omega, 0 , 0 ,k)$; note
that the term proportional to $k^\mu k^\nu$ in $I^{\mu\nu}$ does not
contribute due to the Ward identity.

\subsection{\texorpdfstring{\boldmath$e^+ e^-$}{ee} annihilation to dark states}

 Here we consider the process
$e^-(p_1) + e^+(p_2) \to \chi (p_\chi) + \bar\chi(p_{\bar\chi}) $.  By setting
$p_i = (E_i , \vec{p_i})$ and $k = p_1+p_2$, one can define the cross sections\footnote{We emphasize that this leads to a Lorentz-invariant total cross section up to the thermal mass of photons, which is convenient for the phase space integral.}
in terms of the squared matrix element for annihilation
$|\mathcal{M}_{\mathrm{ann}}|^2 $,
\begin{align}
    \sigma &=   \int \dfrac{d\Pi_{i=\chi,{\bar\chi}}}{4E_1 E_2 v_M}  (2\pi)^4 \delta^4 (k-p_\chi -p_{\bar\chi})
  \dfrac{1}{4} \sum\limits_{\text{spins}} |\mathcal{M}_{\mathrm{ann}}|^2 \notag \\
&= \dfrac{\pi \alpha}{4E_1 E_2 v_M} D_{\mu\nu} D^{*}_{\rho\sigma}   \mathcal{T}_e^{\mu\rho}  I_\chi^{\nu\sigma}\,,
 \label{eq:fullSigmaAnn}
\end{align}
where $\mathcal{T}^{\mu\rho}_e$ reads
\begin{equation}
\label{Eq:Trace}
\mathcal{T}_e^{\mu\rho}  = -2(s g^{\mu\rho} - 2p_1^\mu p_2^\rho -2p_1^\rho p_2 ^\mu), \\
\end{equation}
and $I_\chi^{\nu\sigma} $ is given in \eqref{Eq:PhaseI}; here $s = k^2$. Furthermore,
$v_M$ is the M{\o}ller velocity defined as $v_M = F/(E_1 E_2)$ and the
flux factor $F$ is given by
\begin{equation}
\label{Eq:flux_factor}
F=[(p_1 \cdot p_2)^2 -m_e^4]^{1/2} =\dfrac{1}{2} \sqrt{s(s-4m_e^2)}\,.
\end{equation}
Contracting the Lorentz indices then yields 
\begin{equation}
	 \sigma =  \sigma_T +  \sigma_L\,, 
\end{equation}
where the interference term vanishes in the Coulomb gauge, as can also be seen from \eqref{eq:generalRate}, and the cross section for each polarization mode is given as equations 
(\ref{Eq:Final_ann_cross_T}) and (\ref{Eq:Final_ann_cross_L}) in
the main text.

\subsection{\texorpdfstring{\boldmath$e^- N$}{eN} bremsstrahlung production of dark states}
\label{sec:bremsstr-prod-dark}

The  $2\rightarrow 3$ amplitude squared
$|\mathcal{M}_{2\rightarrow 3}|^2$ can be split into three parts as
\begin{equation}
  \label{eq:MelBrems}
\sum_{\rm spins} \dfrac{|\mathcal{M}_{2\rightarrow 3}|^2}{(4\pi \alpha)^2  g_1 g_2} =  D^{\rho\beta }(	q)D^{\sigma\gamma*}(q) W_{\rho\sigma} L^{\mu\nu}_{\beta\gamma}  \epsilon_\mu^* (k) \epsilon_\nu (k),
\end{equation}
where $q=p_2-p_4$ is the momentum transfer between the initial states,
$L^{\mu\nu}_{\beta\gamma}$ stands for the leptonic part,
$W_{\rho\sigma} $ is the hadronic tensor and $\epsilon_\nu (k)$ is
the polarization vector of the emitted photon of virtual mass
$\sdark = k^2$. 
Detailed forms for $L^{\mu\nu}_{\beta\gamma}$ and $W_{\rho\sigma} $
are given in the App.~A of our previous work~\cite{Chu:2018qrm}.

The $2 \rightarrow 3 $ cross section reads 
\begin{equation}
\sigma_{2 \rightarrow 3} = \dfrac{1 }{4 g_1 g_2 E_1 E_2 v_M}\int d\Pi_{i=3,4,k}  \sum\limits_{\text{spins}} |\mathcal{M}_{2\rightarrow 3}|^2   \,,
\end{equation}
where $v_M$  is the M{\o}ller velocity, as defined in the main text. The phase space integrations, when written in terms of Lorentz invariants
reads,
\begin{equation}
\begin{split}
\sigma_{2\rightarrow 3} = &\dfrac{1}{32(2\pi)^4 E_1 E_2 v_M} \int ds_4 \int dt_1 \dfrac{1}{\sqrt{\lambda (s_4, m_p^2 , t_1)}} \\ 
&\times \int dt_2 \int dp_{1k} \left| \dfrac{\partial  \phi_4^{R4k}}{\partial  p_{1k}}\right| \dfrac{1}{g_1 g_2} \sum\limits_{\text{spins}} |\mathcal{M}_{2\rightarrow 3}|^2 .
\end{split}
\end{equation}
Here, $s= (p_1 + p_2)^2$, $t_1 \equiv (p_1 -p_3)^2$,
$t_2 \equiv (p_2 - p_4)^2 = q^2$, $s_4 \equiv (p_4 + k)^2$, $p_{1k} = p_1 \cdot k$ and $\phi_4^{R4k}$
is the azimuthal angle between $p_4$ and $k$ in their center of mass
frame; $\lambda ( a^2, b^2 ,c^2)$ is the K\"{a}ll\'{e}n function.  The
Jacobian $|\partial \phi_4^{R4k} /\partial p_{1k}|$ transforms the
variable $\phi_4^{R4k}$ to the Lorentz invariant variable $p_{1k}$.
The integration boundary of $s_4$ is given by,
\begin{equation}
 (m_N + \sqrt{\sdark})^2 \leq s_4 \leq   (\sqrt{s} - m_e)^2 ,
\end{equation}
and the boundaries of $t_1$ and $t_2$ are given by
\begin{equation}
\begin{split}
& t_1^\pm = 2m_e^2 - \dfrac{1}{2s} \left[ (s+m_e^2 - m_N^2)(s + m_e^2 -s_4)  \right.\\ 
& \quad\quad\,\, \left. \mp \lambda (s, m_e^2 , m_N^2)^{1/2} \lambda (s , m_e^2 , s_4)^{1/2} \right], \\
& t_2^\pm = 2 m_N^2 - \dfrac{1}{2s_4} \left[ (s_4 +m_N^2 -t_1) (s_4 +m_N^2- \sdark ) \right. \\ 
&  \quad\quad\,\, \left. \mp \lambda (s_4 , m_N^2 ,t_1)^{1/2} \lambda(s_4, m_N^2 ,\sdark)^{1/2} \right].
\end{split}
\end{equation}
The physical region for  $p_{1k}$ is expressed by $n\times n$ asymmetric and symmetric Gram determinants, $G_n$ and $\Delta_n$. It reads,
\begin{equation}
\begin{split}
p_{1k}^\pm &= \dfrac{(p_1 \cdot p_2)G_2 (p_2 , \sqrt{t_1} ; \sqrt{t_1}, k)}{- \Delta_2(p_2, \sqrt{t_1})}\\ & \quad\, -
\dfrac{ (\sqrt{t_1}\cdot p_1) G_2 (p_2, \sqrt{t_1}; p_2, k)}{- \Delta_2(p_2, \sqrt{t_1})}
 \\ & \quad\, \pm \dfrac{\sqrt{\Delta_3 (p_2,\sqrt{t_1},p_1) \Delta_3 (p_2 ,\sqrt{t_1} , k)}}{-\Delta_2(p_2,\sqrt{t_1})},
\end{split}
\end{equation}
and the Jacobian $|\partial \phi_4^{R4k} /\partial p_{1k}|$ reads
\begin{equation}
\left| \dfrac{\partial  \phi_4^{R4k}}{\partial  p_{1k}} \right| =- \dfrac{\sqrt{-\Delta_2(p_2,\sqrt{t_1})}}{\sqrt{-\Delta_4 (p_2 ,\sqrt{t_1} , p_1, k)}}.
\end{equation}

Putting everything together, the full $2\rightarrow 4$ cross section
is given by
\begin{equation}
\sigma_{2\rightarrow 4} = \int d\sdark\,  \sigma_{2\rightarrow 3} (\sdark ) \dfrac{f(\sdark)}{16\pi^2 \sdark^2} \sqrt{1-\dfrac{4m_\chi^2}{\sdark}}.
\end{equation}
The integration boundaries of $\sdark$ are given by
\begin{equation} \label{eqn:boundaries_schi}
4m_\chi^2 \leq \sdark \leq (\sqrt{s}-m_e-m_N)^2.
\end{equation}

\section{Soft-photon approximation for bremsstrahlung}
\label{sec:soft-phot-appr}

Here we discuss the soft-photon approximation for bremsstrahlung, its regime of validity and explain where it fails in calculating the $2\to 4$ cross section.
In the soft limit, that is, if the emitted photon
energy is small compared to the available kinetic energy $\omega \ll E_\text{kin}$, the $2\to 3$ cross section can be factorized into an elastic scattering and an emission part\footnote{The emission part here describes the emission off one of the particles. If both particles can emit photons, the emission part has to be adjusted correspondingly.},
\begin{equation} \label{eqn:dsigma_soft}
d\sigma_{2\rightarrow 3}^\text{\tiny soft} = d\sigma_{2\rightarrow 2} \int \dfrac{d^3 k}{(2\pi)^3 2 \omega} 4\pi \alpha \left| \dfrac{p_3 \cdot \epsilon^*}{p_3 \cdot k} - \dfrac{p_1 \cdot \epsilon^*}{p_1 \cdot k} \right|^2,
\end{equation}
where $ \omega^2  = |\vec{k}|^2 +\sdark$ and $\epsilon_\mu $ is the polarization vector of the emitted photon. In this approximation, a simple form of the differential $2\to 3$ cross section can be obtained in the non-relativistic and ultra-relativistic limit respectively \cite{berestetskii1982quantum},
\begin{align} \label{eqn:soft_limit}
	\omega \frac{d\sigma_{2 \rightarrow 3}^\text{\tiny soft}}{d\omega} = 
	\begin{cases}
		\frac{16}{3} \frac{\alpha^3}{\mu^2 v^2} \ln\left[\frac{1+\sqrt{1-\omega/E_\text{kin}}}{1-\sqrt{1-\omega/E_\text{kin}}}\right] \\
	\frac{4 \alpha^3}{\mu^2} \frac{E'}{E}\left(\frac{E}{E'}+\frac{E'}{E}-\frac{2}{3}\right)\left(\ln\frac{E^2 E'}{\mu^2 \omega}-\frac{1}{2}\right) 
	\end{cases} 
	\end{align}
        where $\mu$ is the reduced mass, $v$ is the relative velocity
        and $E$ ($E'$) is the initial (final) total CM energy of the
        colliding particles.  Even though in deriving the expression in
        Eq.~\eqref{eqn:soft_limit} we have assumed
        $\sqrt{\sdark} \ll E_\text{kin}$ and
        $\omega \ll E_\text{kin}$, integrating \eqref{eqn:soft_limit}
        over $\omega$ in the region
\begin{align}
	\sqrt{\sdark}<\omega<E_\text{kin}
\end{align}
still gives a very good approximation to the full cross section.
Obviously, the approximation breaks down if $\sqrt{\sdark} \sim E_\text{kin}$ where the integration region of $\omega$ gets very small and the integral is dominated by large emission energies.

To obtain the $2\to 4$ cross section from the $2\to 3$ cross section in Eq.~\eqref{Eq:2to4to3}, $\sigma_{2\rightarrow 3}^\text{\tiny soft}$ gets multiplied by the factors in Eq.~\eqref{eqn:f} corresponding to the EM form factor interactions. This leads to the following parametric dependence on $\sdark$,
\begin{align} \label{eqn:soft_cs_models}
	d\sigma_{2\rightarrow 4}^\text{\tiny soft} \propto 
	\begin{cases}
          d\sigma_{2\rightarrow 3}^\text{\tiny soft} \; d\sdark/\sdark &\text{(dim-4)}, \\
          d\sigma_{2\rightarrow 3}^\text{\tiny soft} \; d\sdark &\text{(dim-5)}, \\
          d\sigma_{2\rightarrow 3}^\text{\tiny soft} \; d\sdark\,
          \sdark &\text{(dim-6)},
	\end{cases}
\end{align}
that is, for dim-4 operators, like millicharged states, the $2\to 4$
cross section is dominated by small $\sdark$, whereas for higher
dimensional operators, the expression is UV biased. Hence, the main
contribution to the integral comes from $\sdark$-values for which the
soft approximation breaks down as one probes the kinematic endpoint
region.  It turns out that in the non-relativistic regime and for
$m_\chi+m_{\bar \chi} \ll E_\text{kin}$, using
Eq.~\eqref{eqn:soft_limit} reproduces the exact $2\to 4$ cross section
up to a factor 2 or 3. However, for relativistic particles, the error
at $\sqrt{\sdark}\sim E_\text{kin}$ gets larger. Due to the
$\sdark$-dependence in Eq.~\eqref{eqn:soft_cs_models}, this still
results in a decent description of millicharged $\chi\bar\chi$
emission, but produces errors of several orders of magnitude in the
relativistic regime for the EM form factors considered in this paper.

Equation~\eqref{eqn:dsigma_soft} can be further
simplified by separating the phase space. This is possible, if the
elastic scattering cross section is insensitive to an angular cut-off
in the forward or backward direction, \textit{e.g.}~if the interaction
is mediated by a massive particle such as the pion in $np$
scattering.%
\footnote{For $ep$ scattering, on the other hand, the phase space
  separation is not possible, since the elastic cross section is
  forward divergent. In these cases, 3-body kinematics is required.}
Then,
\begin{equation} \label{eqn:dsigma_soft_sep}
\sigma_{2\rightarrow 3}^\text{\tiny soft} = \sigma_{2\rightarrow 2}^{\text{T}} \, \mathcal{I}(\sdark) \,,
\end{equation}
where $\sigma_{2\rightarrow 2}^{\text{T}}$ is the transport cross section
\begin{align}
	\sigma_{2\rightarrow 2}^{\text{T}} = \int_{-1}^1 d\cos \theta \frac{d\sigma_{2\rightarrow 2}}{ d\cos\theta} (1-\cos\theta) \,,
\end{align}
and the emission piece $\mathcal{I}(\sdark)$ is obtained by executing the integral in Eq.~\eqref{eqn:dsigma_soft}
\begin{align} \label{eqn:emission_piece_sep}
	\mathcal{I}(\sdark) &= \frac{1}{1-\cos\theta}\int \dfrac{d^3 k}{(2\pi)^3 2 \omega} 4\pi \alpha \left| \dfrac{p_3 \cdot \epsilon^*}{p_3 \cdot k} - \dfrac{p_1 \cdot \epsilon^*}{p_1 \cdot k} \right|^2 \notag \\
	&= \frac{\alpha}{3\pi} \int_{\sqrt{\sdark}}^{E_\text{kin}} d\omega \frac{\sqrt{\omega^2-\sdark}\left(\sdark/2+\omega^2\right)}{\omega^4} \,.
\end{align}
In the first line, we have divided by $(1-\cos\theta)$ to cancel the $\theta$-dependent part in the emission piece, which we have absorbed into the elastic cross section.
In Sec.~\ref{sec:nucleon-bremsstrahlung} we make use of this factorization in calculating the energy loss rate for neutron-proton scattering in PNS. For that, the integral in Eq.~\eqref{eqn:emission_piece_sep} is weighted with $\omega$ to obtain
\begin{align}
\begin{split}
	\mathcal{I}_\omega(\sdark)	
	&= \frac{\alpha}{3\pi} \int_{\sqrt{\sdark}}^{E_\text{kin}} d\omega \frac{\sqrt{\omega^2-\sdark}\left(\sdark/2+\omega^2\right)}{\omega^3} \\
	&=\frac{\alpha E_\text{kin}}{3\pi} \left[
	\sqrt{1-x^2}(4-x^2)-3x \arccos(x)
	\right] 
\end{split}\label{eqn:emission_piece}
\end{align}
with $x=\sqrt{\sdark}/E_\text{kin}$,
which is in agreement with the findings of Ref.~\cite{Rrapaj:2015wgs}.

\bibliography{refs}

\end{document}